\documentclass[aps,prx,10pt,twocolumn,superscriptaddress,longbibliography]{revtex4-2}

\usepackage{graphicx}
\usepackage{amssymb,amsmath,amsfonts,gensymb}
\usepackage{bm,hyperref}
\usepackage{color}
\usepackage{ulem}
\usepackage{bbm}
\usepackage{subfigure}
\usepackage{dcolumn}
\usepackage{mathtools}
\usepackage{pifont}

\newcommand{\ket}[1]{\vert #1 \rangle}
\newcommand{\bra}[1]{\langle #1 \vert}

\def\I{\uppercase\expandafter{\romannumeral 1}}
\def\II{\uppercase\expandafter{\romannumeral 2}}
\def\III{{\uppercase\expandafter{\romannumeral 3}}}
\def\IV{{\uppercase\expandafter{\romannumeral 4}}}
\def\V{{\uppercase\expandafter{\romannumeral 5}}}
\def\VI{{\uppercase\expandafter{\romannumeral 6}}}
\def\VII{{\uppercase\expandafter{\romannumeral 7}}}
\def\i{\lowercase\expandafter{\romannumeral 1}}
\def\ii{\lowercase\expandafter{\romannumeral 2}}
\def\iii{{\lowercase\expandafter{\romannumeral 3}}}
\def\iv{{\lowercase\expandafter{\romannumeral 4}}}
\def\v{{\lowercase\expandafter{\romannumeral 5}}}
\def\vi{{\lowercase\expandafter{\romannumeral 6}}}
\def\vii{{\lowercase\expandafter{\romannumeral 7}}}

\newcommand{\bk}{\mathbf{k}}
\newcommand{\btk}{\widetilde{\mathbf{k}}}
\newcommand{\btq}{\widetilde{\mathbf{q}}}
\newcommand{\br}{\mathbf{r}}

\def\nn{\nonumber \\}

\def\p{\mathbf{p}}

\def\k{\mathbf{k}}
\def\vr{\mathbf{r}}
\def\G{\mathbf{G}}
\def\Q{\mathbf{Q}}
\def\br{\mathbf{r}}
\def\kt{\widetilde{\mathbf{k}}}
\def\q{\mathbf{q}}

\def\hc{\hat{c}}
\def\hcd{\hat{c}^{\dagger}}

\begin{document}

\title{A Numerical Perspective on Moir\'e Superlattices: From Single-Particle Properties to Many-Body Physics}

\author{Xin Lu}
\email{lvxin@shanghaitech.edu.cn}
\affiliation{School of Physical Science and Technology, ShanghaiTech Laboratory for Topological Physics, State Key Laboratory of Quantum Functional Materials, ShanghaiTech University, Shanghai 201210, China}

\author{Bo Xie}
\affiliation{School of Physical Science and Technology, ShanghaiTech Laboratory for Topological Physics, State Key Laboratory of Quantum Functional Materials, ShanghaiTech University, Shanghai 201210, China}

\author{Jianpeng Liu}
\email{liujp@shanghaitech.edu.cn}
\affiliation{School of Physical Science and Technology, ShanghaiTech Laboratory for Topological Physics, State Key Laboratory of Quantum Functional Materials, ShanghaiTech University, Shanghai 201210, China}
\affiliation{Liaoning Academy of Materials, Shenyang 110167, China}
	
\bibliographystyle{apsrev4-2}

\begin{abstract} 
    Moir\'e superlattices in two-dimensional materials provide a versatile platform to explore strongly correlated and topological phases. This work presents a practical theoretical workflow for studying the correlated and topological states in moir\'e systems, combining continuum modeling, Hartree-Fock mean-field approximations, many-body perturbation theory, and exact diagonalizations. We focus on the numerical implementation of these methods, emphasizing subtleties such as remote band effects, inhomogeneous and dynamical screening, double counting problem, etc., which are often swept under the rug in theoretical works. The workflow enables a systematic investigation of symmetry-breaking ground state properties, quasiparticle excitation properties and fractional Chern insulator phases emerging from moir\'e superlattices, providing insights that are directly relevant to experimental observations. By bridging technical details and physical interpretations, this work aims to guide both theorists and experimentalists in understanding and predicting correlated phenomena in moir\'e materials.
\end{abstract} 

\maketitle

\section{Introduction}
One of the most fascinating aspects of condensed matter physics is the unexpectedly rich many-body phenomena that emerge from the assembly of individual atoms into a lattice, embodying Anderson's famous principle that ``more is different'' \cite{more-is-different}. Before the discovery of moir\'e system, strongly correlated phenomena such as unconventional superconductivity \cite{lee-rmp06,keimer-nature-2015}, Mott insulators \cite{fazekas-book,mit-rmp-1998}, and Kondo effect \cite{Hewson_1993} were typically observed and studied in materials  such as transition metal oxides and rare earth components, often referred to as strongly correlated materials. To understand the peculiar properties of strongly correlated materials, one typically needs ``tuning knobs'' to change carrier density, band gaps, bandwidths, band dispersions, etc., and observe how the ground states and excitations would evolve under the tuning. However, it has been a critical challenge to alter the intrinsic properties of traditional strongly correlated materials in an ``in situ'' manner, without introducing various undesirable extrinsic effects. For example, to vary carrier density, substitutional impurity atoms have to be introduced, which would unavoidably introduce disorders. To tune the ratio between kinetic energy and interaction energy, one may need to apply pressure to the material, which certainly modifies the electronic bandwidth but also leads to significant structural distortions and even structural transitions. In such a context, reviving these strongly correlated phases in materials traditionally regarded as weakly correlated, such as two-dimensional (2D) graphene \cite{graphene-rmp-09} and transition metal dichalcogenides (TMDs) \cite{tmd-nrm-2017}, represents a significant breakthrough. Most importantly, due to the 2D nature of these layered van der Waals materials, their electronic properties can be easily tuned by electric gate voltages.

The key to inducing strong correlations in these van der Waals 2D systems lies in twisting the layers at small angles, which results in the formation of flat bands \cite{trambly-nanoletter10,suarez-tbgflatband-prb-2010,macdonald-pnas11,wu-TI-twisted-tmd-prl19}. 
The moir\'e pattern, which arises from the atomic lattice mismatch due to twisting, induces a periodic  potential that defines a mini Brillouin zone, known as the moir\'e Brillouin zone. This moir\'e potential is crucial for the emergence of topological flat bands. It not only folds the atomic bands into the tiny moir\' Brillouin zone, gapping out the degeneracy  points due to band folding, thereby reducing the bandwidth; but more importantly, it induces a pseudo gauge field that generates an effective periodic magnetic field \cite{jpliu-prb19,moralesduran-maTMD-prl-2024}. This pseudo magnetic field imparts non-trivial topological wave functions to the flat bands \cite{origin-magic-angle-prl19,song-tbg-prl19,yang-tbg-prx19,jpliu-prb19,wu-TI-twisted-tmd-prl19,jpliu-prx19,song-tbg-ii-prb21}, making them exactly flat in certain limits as in the case of magic-angle twisted bilayer graphene, similar to Landau levels \cite{origin-magic-angle-prl19}. This makes moir\'e systems an ideal platform for investigating the interplay between topology and $e$-$e$ interactions. Compelling examples are featured by the emergence of fractional Chern insulator (FCI) \cite{fci-prx11,sheng-fci-nc11,murdy-fci-prl11,wen-kagome-prl11,sarma-flatchern-prl11,qi-fqah-prl11}, lattice realization of fractional quantum Hall effect without an actual magnetic field, in twisted TMDs (tTMDs) \cite{fqah-nature23,fqah-prx23,fqah-optics-xu-nature23,fqah-mak-nature23} and hexagonal boron nitride-rhombohedral $n$-layer graphene (hBN-R$n$G) moir\'e systems \cite{fqah-julong-nature24,fqah-luxb-natmat25,fqah-julong-nature25}.

In the field of moir\'e systems, both experimental and theoretical efforts are highly active and strongly synergistic since the proposition of continuum model for twisted bilayer graphene (TBG) \cite{santos-tbg-prl07,macdonald-pnas11}. Nevertheless, if there is one experimental breakthrough that sparked widespread enthusiasm among physicists for the study of moir\'e systems, it would undoubtedly be the observation of superconductivity in magic-angle TBG in 2018 \cite{cao-nature18-supercond}. This discovery highlights how strongly correlated phenomena could be realized in moir\'e systems. Although theorists had predicted the emergence of flat bands in moir\'e systems such as magic-angle TBG \cite{macdonald-pnas11} several years before, a variety of novel correlated phases, first revealed by experiments, go beyond theoretical predictions and several of them still remain not well understood. There are several reasons for this. On the experimental side, it reflects the long-standing endeavors of experimentalists in developing state-of-the-art device fabrication techniques \cite{dean-hBN_sub-natnano-2010}, which have been advanced over more than a decade since the discovery of monolayer graphene in 2004 \cite{graphene-science-04}, well before moir\'e systems were even identified. On the theoretical side, the difficulties stem not only from the notorious challenges of many-body theory itself, but also from additional complications unique to moir\'e systems. First, the large moir\'e unit-cell typically contains thousands of atoms at small twist angles. This poses challenges to existing density functional theory (DFT)-based numerical packages designed for small unit-cells, such as DFT+$U$, DFT with dynamical mean-field theory (DMFT) \cite{georges-dmft-rmp-1996}, etc., to handle moir\'e systems effectively and economically. Moreover, due to the topologically non-trivial nature of the low-energy electrons in moir\'e systems, it is also hard to construct effective lattice models on the moir\'e length scale. As a result, theoretical frameworks tailored for moir\'e systems call for a systematic generalization of existing methods, a redesign of computational workflows, and a re-optimization of numerical routines to wrestle with the challenges posed by the large number of atoms per moir\'e unit-cell and the system's non-trivial topology.

A detour taken by theorist to study the low-energy physics in moir\'e systems involves expanding the low-energy moir\'e bands in a plane-wave basis near high-symmetry points in the Brillouin zone, known as continuum model \cite{santos-tbg-prl07,shallcross-tbg-prb-2010,macdonald-pnas11,castro-neto-prb12,wu-TI-twisted-tmd-prl19}. The most successful achievement of this approach is perhaps the prediction of a series of magic angles in TBG \cite{macdonald-pnas11,suarez-tbgflatband-prb-2010}, around which superconductivity was observed. While there have been attempts to derive effective models within an even lower energy window (around the flat bands) based on the continuum model \cite{song-heavyfermion-prl22,po-tbg-prb19}, the latter still remains the most widely accepted non-interacting model because it respects all the symmetries of the system and has already yielded numerous significant results that help explain and predict experimental observations in a computationally efficient way.

Compared to DFT-based methods, the primary advantages of the continuum model approach lie in its ability to treat $e$-$e$ interactions and topology in moir\'e systems in a highly efficient yet accurate manner. From a technical perspective, the large number of orbitals in a moir\'e unit-cell makes computations for interacting physics highly intractable using existing DFT-based numerical routines such as many-body perturbation theory \cite{rubio-rmp-2002,giustino-rmp-2017,coleman-book} and DMFT \cite{georges-dmft-rmp-1996}. One might consider using an effective Wannier-function-based approach (on the moir\'e length scale) to capture the low-energy physics within DFT-based methods, as is typically required in DMFT \cite{georges-dmft-rmp-1996} and DFT+$U$ \cite{liechtenstein-prb-1995} calculations. However, in many cases this is forbidden due to topological obstructions, since the low-energy flat bands are usually topologically non-trivial. In contrast, the continuum model, which focuses on low-energy physics, significantly reduces the number of degrees of freedom, making the numerical implementation both feasible and efficient while still accurately capturing the essential physical features. Therefore, given current computational limitations, these technical difficulties already make the continuum model approach an indispensable tool for studying the interplay between correlations and topology in moir\'e systems. 

On a more fundamental level, standard DFT poorly captures strong correlations, which are prevalent in moir\'e systems due to the flatness of the moir\'e bands. In contrast, the continuum model approach has been proven effective in explaining various experimental observations in the strong-interaction regime. For example, moir\'e systems exhibit non-rigid-band behaviors that depend on electron density, as clearly demonstrated by the ``cascade transition behavior'' \cite{Yazdani-cascade-nature20,Ilani-cascade-nature20} in the carrier-density dependence of compressibility measurements, particularly in the flat-band limit where interactions dominate. The symmetries of the many-body ground states are usually completely different for different carrier densities or moir\'e band filling factors. For example, in magic-angle TBG aligned with hBN, the ground state at charge neutrality is a trivial insulator \cite{young-tbg-science-2019,lu-nsr-2025}; while that at filling 3 becomes a Chern insulator spontaneously breaking time-reversal symmetry \cite{young-tbg-science-2019,lu-nsr-2025}. As a result, many-body perturbation theory built on DFT can fail if the DFT-derived ground state (typically assuming charge neutrality) is not adiabatically connected to the experimentally observed correlated and/or topological phase. This limitation is even more severe for FCI, a topologically nontrivial strongly correlated phase exhibiting the fractional quantum anomalous Hall effect, which cannot be captured by either of the aforementioned techniques. Fortunately, these strongly correlated phenomena can be effectively handled by methods developed based on the continuum model, as demonstrated in this work, highlighting the necessity of the continuum model approach from a physical perspective.

The challenge in formulating a theoretical methodology with predictive power within the continuum model lies both in the non-interacting and interacting considerations. The former stems from the large number of atoms in the moir\'e unit-cell, making the study of lattice relaxation effects extremely computationally demanding. Lattice relaxations, however, are particularly strong, especially at small twist angles. For example, in MoTe$_2$, both the in-plane and out-of-plane strain amplitudes were found to increase as the twist angle decreases. The maximum in-plane strain amplitude reached approximately 0.4\;\AA\, at $2.88^\circ$, which is substantial when compared to the MoTe$_2$ lattice constant of 3.52\;\AA\,\cite{mao-TMDrelax-comphy-2024}. The strong relaxation effects typically tend to minimize the area of one type of stacking relative to another in order to save interlayer binding energy, and there are also other distortion-related effects \cite{koshino-tbg-prb17,koshino-prx18,kazmierczak2021strain,cantele-prr-2020}. These effects are crucial for establishing a reliable non-interacting starting point for subsequent studies of interacting physics. For instance, the moir\'e phonon structure \cite{angeli-tbg-prx19,liu-phonon-nano22,xie_phonon_prb23,kaxiras-phonon-prb} is important for understanding the superconductivity and other phenomena linked to electron-phonon coupling \cite{eliel2018intralayer,chen-replica-nature24}. Another example is the overestimation of the magic angle for twisted MoTe$_2$ \cite{moralesduran-maTMD-prl-2024}, where flat bands were initially believed to emerge around smaller twist angles $\lessapprox 1.5^{\circ}$, while experimentally fractional quantum anomalous Hall effects (in twisted MoTe$_2$) \cite{fqah-nature23,fqah-prx23} have been observed under more than twice larger angles. This error is believed to originate from inaccurate material-related parameters in the continuum model and the insufficient consideration of lattice relaxation effects. It is also important to note that these parameters determined by experiments are typically renormalized by Coulomb interactions (particularly the nonlocal exchange effects \cite{kang-rg-prl20,guo-HFfci-prb-2024}), so they cannot be used directly in the non-interacting continuum models, otherwise leading to double counting of $e$-$e$ interaction effects. These parameters would be better extracted from first principles DFT calculations that account for lattice relaxations, but the nonlocal exchange effects are not included. Therefore, obtaining an accurate non-interacting continuum model, which serves as the foundation for subsequent interacting theory, depends on the support of DFT-based methods. In this work, we will first discuss how to incorporate relaxation effects into continuum models with limited computational resources.

Once a plausible non-interacting continuum model is established, the treatment of interactions presents a second major challenge, which is notoriously difficult in general. For example, there is still non consensus on the mechanism of superconductivity observed in magic-angle TBG. Moreover, implementing conventional methods such as Hartree-Fock mean-field approximations, random phase approximations, and $GW$ approximations involves additional subtleties. For example, Hartree-Fock approximations typically require truncating high-energy bands, projecting the Coulomb interactions onto a low-energy subspace, and limiting self-consistent calculations to this active space. However, there are several methods for doing the projection, each of which handles interactions between electrons in the truncated high-energy bands and those in the projected low-energy bands in different ways \cite{guo-HFfci-prb-2024,zhou-R5Gfci-prl-2024,dong-R5Gfci-prl-2024,dong-R5GAHC-prl-2024,kwan-RnGHF-prb-2025}. None of these methods can be definitively said to be superior to the others. This difficulty arises because the gaps between moir\'e bands are often small compared to the strength of the interactions, making it challenging to reach a consensus on the appropriate cut-off energy, other than the plane-wave cut-off, which defines the ultraviolet boundary of continuum theory. A similar issue leads to the double-counting problem and the multi-band effect when using exact diagonalization to study FCI states in moir\'e systems \cite{guo-HFfci-prb-2024,zhou-R5Gfci-prl-2024,dong-R5Gfci-prl-2024,dong-R5GAHC-prl-2024,yu-R5GED-prb-2025,huang-fqahdoublecounting-prb-2024}. We will address these subtleties in this work.

\begin{figure}
    \centering
    \includegraphics[width=0.45\textwidth]{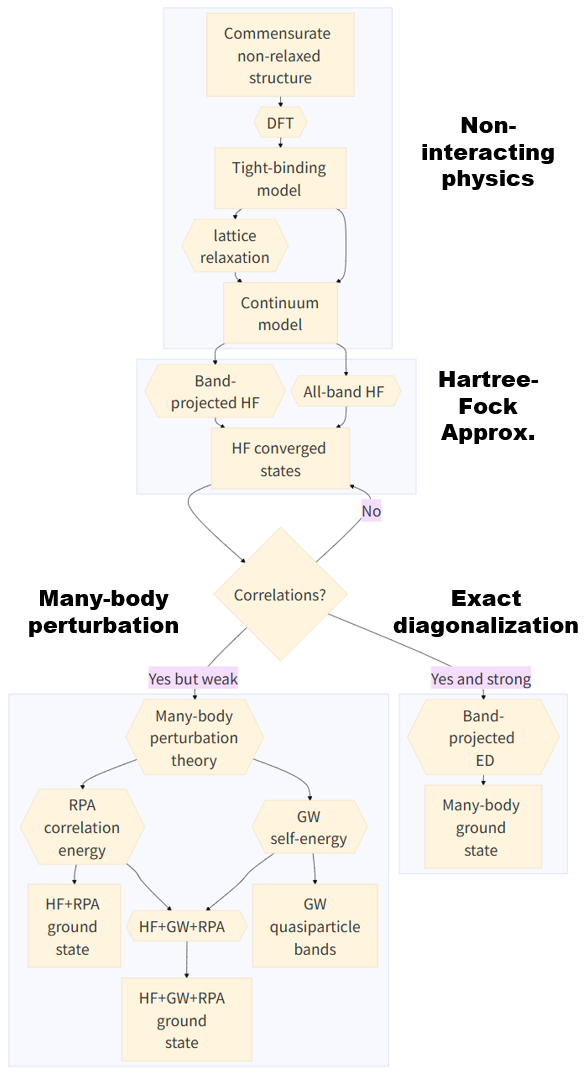}
    \caption{Workflow to study the non-interacting and interacting effects in moir\'e systems. The large shaded rectangular frames with bold title nearby denote different computational levels, which corresponds one-to-one the sections in the present work. Small rectangular boxes in each frame denote the input and output data for methods, which are represented by elongated hexagonal boxes. Arrows indicate data dependencies.}
    \label{fig:workflow}
\end{figure}

The purpose of this work is to present a theoretical workflow based on the continuum model, using numerical techniques that are commonly regarded as physically relevant in the community due to already gained success. As summarized by the flow chart in Fig.~\ref{fig:workflow}, this workflow begins with studying how individual atoms form moir\'e superlattices in the presence of lattice relaxations to get non-interacting parameters in continuum models, which can reproduce the DFT results at low energy to remarkable accuracy. Based on the obtained models, we apply Hartree-Fock (HF) mean-field approximations to explain various spontaneous symmetry-breaking states driven by $e$-$e$ interactions in moir\'e systems, which usually emerge at partial integer fillings \cite{kang-tbg-prl19,zaletel-tbg-hf-prx20,Lian-tbg4-arxiv20}. This is followed by the application of many-body perturbation theory to determine the actual ground states, when different states compete, using the random phase approximation (RPA) for correlation energy, and get the quasiparticle band structures within $GW$ approximations. One can even perform a more accurate calculation, which is to calculate the RPA correlation energy based $GW$ quasiparticles. Such ``$GW$+RPA'' method turns out to be much more accurate than HF+RPA method when treating Wigner-crystal transition in two-dimensional electron gas model \cite{guo-AHC-arxiv-2025}. For the study of FCI states and their competing phases in fractionally filled topological flat bands, we turn to exact diagonalization techniques, which are widely used in the quantum Hall problem. Throughout this work, we focus on the implementation of these methods mostly from our own perspective, which have already proven promising in explaining various experimental results such as the competition between different isospin polarized states \cite{Liu-magnetic-vp-tdbg-nc22,zhang-tbmg-prl22,Li-mao-tbmg-exp-nc22,liu-1storderTDBG-prx-2023,jpliu-tbghf-prb21,zhang-liu-tbg-prl22,efetov-nature19,young-tbg-science19,sharpe-science-19,efetov-nature20} and the emergence of FCIs in hBN-R5G heterostructure \cite{guo-HFfci-prb-2024,dong-R5GAHC-prl-2024, senthil-fqah-prl24, zhang-fqah-prl24, fqah-julong-nature24,fqah-luxb-natmat25}. We also provide a brief overview of the implementations of the same methods by other groups and from other perspectives, if they differ, for comparison. The goal of this work is not to provide an extensive discussion of the physics of moir\'e systems, which would require an entire monograph. Instead, we offer a numerical perspective focusing on the technical aspects of the numerical implementations, which are often swept under the rug in theoretical works. This work aims to help readers to understand what the theoretical outcomes using these methods truly mean, bridging the gap between experimentalists and theorists, and discussing both their strengths and limitations. Additionally, we hope to introduce theorists unfamiliar with moir\'e systems to this field, encouraging them to further enrich the theoretical toolbox for studying moir\'e systems.

The work is organized as follows. In Sec.~\II, we discuss how lattice relaxation is incorporated into the construction of the continuum model. Specifically, we explain how we overcome the challenges in \textit{ab initio} calculations when determining model parameters at small twist angles. In Sec.~\III, we demonstrate how to implement HF approximations in the continuum model, emphasizing the importance of the remote band effects. In Sec.~\IV, we present many-body perturbation theory for moir\'e systems based on the previously obtained HF states, a topic that has rarely been explored. In Sec.~\V, we describe how exact diagonalization techniques are used to study FCI states in moir\'e systems, highlighting several numerical subtleties tied to moir\'e systems. Finally, we summarize the overall workflow for moir\'e systems and discuss the next steps for further development. We refer the readers to the corresponding papers that we provide at appropriate places for the detailed derivations of the formula given in the work. 

\section{Continuum model considering lattice relaxation}
\label{sec:contmodel}
To study the physics of moir\'e superlattice systems, researchers have developed a variety of theoretical models. These models span from \textit{ab initio} approaches that rigorously account for atomic-scale details to simplified effective theories that focus on low-energy physics. DFT stands as a cornerstone \textit{ab initio} method for investigating electronic properties, as it directly links the lattice structure of a material to its electronic band structure by solving the Kohn-Sham equations \cite{kohn-sham}. However, DFT suffers from a critical limitation: its computational cost scales polynomially with the number of atoms (typically $\mathcal{O}(N^{3})$ for standard iterative solvers, where $N$ is the number of electrons). For moir\'e superlattices, which typically contain thousands of atoms per unit-cell, this scaling renders DFT computationally intractable. The memory and time requirements to converge Kohn-Sham orbitals for such large systems become prohibitive, making DFT unsuitable for studying the moir\'e superlattice systems.

An alternative atomistic approach is the tight-binding (TB) model, which reduces computational complexity by focusing on a limited set of orbitals per atom. TB models parameterize electron hopping between neighboring atoms using empirical or DFT-derived hopping amplitudes, which builds up an explicit link between the Hamiltonian and the microscopic hopping based on the lattice geometry. Additionally, TB models reduce the computational cost by reducing the size of Hilbert space (e.g., compared to plane-wave basis DFT calculations) and by avoiding the iteration algorithm to solve the Kohn-Sham equation self-consistently. However, even TB models face severe challenges in moir\'e superlattices. Taking magic-angle ($\theta\!=\!1.05^{\circ}$) twisted bilayer graphene (MATBG) as an example, if only the $p_{z}$ orbitals (the primary contributors to $\pi$-bonding and low-energy electronic states) of each carbon atom is taken into account, the number of orbitals exceeds 13 000. Such a large number of single-particle orbitals brings significant problem when incorporating $e$-$e$ interactions. Crucially, the essential low-energy physics of moir\'e superlattices is not governed by atomic-scale details (e.g., individual C-C bond lengths or orbital hybridization) but by the low-energy subspace. This insight justifies the shift from atomistic models to effective continuum models, which focus on the long wavelength (small momentum) electronic degrees of freedom. This simplification makes it feasible to incorporate $e$-$e$ interactions, using methods like mean-field theory, or exact diagonalization, to study their interplay with band topology.

In this section, we discuss the construction of the effective continuum model, a widely used method to describe the low-energy electronic behavior in moir\'e superlattices. We begin by establishing the lattice and reciprocal lattice geometry of such systems, using twisted bilayer structures as an example. For the $l$-th layer, we denote the lattice vectors as $\textbf{a}_{i}^{(l)}$, with corresponding reciprocal lattice vectors $\textbf{a}_{i}^{*,(l)}$, where $i\!=\!1,2$. Each unit-cell contains multiple sublattices (e.g., the $A$ and $B$ sublattices of graphene), with position of the $X$-th sublattice given by $\bm{\tau}_{X}$. When the two layers are misaligned (via twist) or exhibit a small lattice mismatch, the reciprocal lattice vectors of the two layers differ slightly. This mismatch gives rise to the moir\'e reciprocal lattice vector, defined as $\textbf{G}_{i}^{M}\!=\!\textbf{a}_{i}^{*,(1)}-\textbf{a}_{i}^{*,(2)}$. The corresponding real space lattice vectors of the moir\'e superlattice are denoted as $\textbf{L}_{i}^{M}$, which form a commensurate lattice with moir\'e translation symmetry. In Fig.~\ref{fig:structure_band}(a), we show the lattice structure and the corresponding reciprocal lattice of a typical moir\'e superlattice. The lattice constant is denoted as $L_{s}$. In Fig.~\ref{fig:structure_band}(b), AA/AB/BA represents local high symmetry stacking pattern and SP stands for saddle point.

\subsection{Continuum model for twisted bilayer graphene}
The low-energy physics of TBG can be described by such a continuum model, commonly known as Bistritzer-MacDonald model (hereafter referred to as the ``BM model'') \cite{macdonald-pnas11}. This model focuses on the electronic states near the Dirac points of the graphene layers, where the linear dispersion of graphene is preserved. The single particle Hamiltonian of the BM model for the $\mu\!=\!\pm$ valley is given by:
\begin{align}
\begin{split}
H_{\mu}^{TBG}=\left[\begin{array}{cc}
				\hbar v_{F}(\textbf{k}-\textbf{K}_{\mu}^{(1)})\cdot\sigma^{\mu} & U^{\dagger}_{\mu} \\
				 U_{\mu}& \hbar v_{F}(\textbf{k}-\textbf{K}_{\mu}^{(2)})\cdot\sigma^{\mu}
\end{array}\right],
\end{split}
\end{align}
where $\sigma^{\mu}\!=\![\mu\sigma_{x}, \sigma_{y}]$, $\mu\!=\!\pm$ is the valley index, $\textbf{K}_{\mu}^{(l)}$ denotes the $\mu$ atomic valley from $l$-th layer. $v_{F}$ is the Fermi velocity of the Dirac cone. $U_{\mu}(\textbf{r})$ denotes the moir\'e potential for the $\mu$ valley:
 \begin{align}
\begin{split}
U_{\mu}(\textbf{r})=\left[\begin{array}{cc}
				u\,g_{\mu}(\textbf{r}) & u\,g_{\mu}(\textbf{r}+\textbf{r}_{AB}) \\
				 u\,g_{\mu}(\textbf{r}-\textbf{r}_{AB}) & u\,g_{\mu}(\textbf{r})
\end{array}\right]e^{-\mu\Delta\textbf{K}\cdot\textbf{r}},
\end{split}
\end{align}
where $\textbf{r}_{AB}\!=\![\sqrt{3}L_{s}/3,0]$, and $\Delta\textbf{K}\!=\!\textbf{K}^{(2)}_{+}-\textbf{K}^{(1)}_{+}$. $u$ represents the amplitude of the moir\'e potential. The phase factor $g(\textbf{r})$ is defined as $g_{\mu}(\textbf{r})\!=\!\sum_{j=1}^{3}e^{-i\mu\textbf{q}_{j}\cdot\textbf{r}}$, where $\textbf{q}_{1}\!=\![0,4\pi/3L_{s}]$, $\textbf{q}_{2}\!=\![2\pi/\sqrt{3}L_{s},-2\pi/3L_{s}]$ and $\textbf{q}_{3}\!=\![-2\pi/\sqrt{3}L_{s},-2\pi/3L_{s}]$ are three leading order moir\'e wavevectors with dominant Fourier components of the moir\'e potential. The emergence of the low energy subspace in the moir\'e superlattice can be physically interpreted as a two-step process, as captured by the BM model: first, the energy bands fold into the mini Brillouin zone of the moir\'e superlattice; second, the moir\'e potential hybridizes these folded low energy states. For TBG near magic angle $\theta\!=\!1.05^{\circ}$, this hybridization nearly cancels the kinetic energy of the electrons, resulting in flat bands with bandwidths smaller than $10\,$meV. Notably, the BM model is not merely a qualitative tool: it quantitatively predicts the existence of flat bands at the correct magic angle long before their experimental observation, and it correctly reproduces the topological properties of these flat bands. This success solidifies the BM model as a foundational framework for extending to other moir\'e systems. 

A simplification of the original BM model is its treatment of the moir\'e superlattice as a rigid lattice, ignoring lattice relaxation effects: in real twisted bilayers, atoms alter their positions to minimize total energy, rather than remaining fixed in their pristine lattice sites. To describe such effects in moir\'e superlattices, we introduce the in-plane and out-of-plane lattice distortions, denoted as $\textbf{u}^{(l)}(\textbf{r})$ and $h^{(l)}(\textbf{r})$. We further introduce linear combinations of these distortion fields: $\textbf{u}^{\pm}\!=\!\textbf{u}^{(2)}\pm\textbf{u}^{(1)}$ and $h^{\pm}\!=\!h^{(2)}\pm h^{(1)}$. In Fig.~\ref{fig:structure_band}(b) (from Ref.~\cite{xie_phonon_prb23}), we present the lattice relaxation pattern of TBG with $\theta\!=\!1.05^{\circ}$. We present the real space distribution of the relative in-plane distribution $\textbf{u}^{-}(\textbf{r})$. The colorbar encodes the amplitudes the distortion fields and the vectors shows the directions of the distortion fields. The in-plane relative distortion fields form a rotational field winding around the AA region, decreasing the area of the AA region. We also show the real space distribution of the out-of-plane relative distortion fields $h^{-}(\textbf{r})$ (the interlayer distance between two layers), with the colorbar represents the amplitudes of the distortion fields. The interlayer distance reaches a maximum value at AA point and a minimum value at AB/BA point. For TBG, lattice relaxation has been shown to narrow the bandwidths of the flat bands and opens up a gap between flat bands and remote bands. Another important extension addresses the non-locality of the moir\'e potential. The original BM model assumes a local moir\'e potential. In TBG, non-locality of the moir\'e potential arises due to the low-energy nature of the continuum theory: essentially it is derived from a local, microscopic, atomic-scale Hamiltonian that is projected to the low-energy subspace around the Dirac points of graphene. Such projection leads to nonl-ocality of the potential, which couples electrons across extended regions of the moir\'e unit-cell \cite{Song-tbg2-arxiv20}. In Fig.~\ref{fig:structure_band}(c) (from Ref.\cite{xie_phonon_prb23}), we present the energy band structures of TBG with $\theta\!=\!1.05^{\circ}$. In the upper subfigure, we present the band structures of TBG calculated by the continuum model. The red dashed lines represent the band structures based on the rigid lattice structure, while the blue lines represent the fully relaxed band structures. The lattice relaxation effects would open up a gap between flat bands and remote bands. In the lower subfigure, we present the band structures based on a fully relaxed lattice structure. The red lines represent the band structures calculated by TB model, while the blue lines represents the band structures calculated by the continuum model considering both the lattice relaxation effects and the non-local moir\'e potential. By considering the lattice relaxation effects and the non-local moir\'e potential, we are able to reproduce the energy band structures calculated by TB model with an efficient low-energy continuum model. In other words, both the lattice relaxation effects and the non-local moir\'e potential play important roles in TBG with small twist angle.

\begin{figure*}
\includegraphics[width=0.80\textwidth]{fig/structure_band_new.png}\caption{(a) The lattice structure and reciprocal lattice of a moir\'e superlattice. Moir\'e superlattice system consists of several layered materials. The lattice mismatch gives rise to an enlarged moir\'e supercell, preserving an additional translational symmetry, and the corresponding mini Brillouin zone. (b) The real space distribution of the lattice relaxation pattern of twisted bilayer graphene with twist angle $\theta\!=\!1.05^{\circ}$. $\textbf{u}^{-}$ represents the relative in-plane distortion. The colorbar encodes the amplitudes of the distortion fields and the vectors show the directions of the distortion fields. $h^{-}$ represents the out-of-plane relative distortions (the interlayer distance), with the colorbar encoding the amplitudes. (c) The band structure of twisted bilayer graphene with twist angle $\theta\!=\!1.05^{\circ}$. The upper sub-figure shows the energy band structures calculated by the continuum model. The red dashed lines represent the band structures based on rigid lattice, while the blue lines represent energy band structure considering the lattice relaxation effects. The lower sub-figure shows the energy band structure based on relaxed lattice structure, the red lines show the band structure calculated by tight binding model, while the blue lines are calculated by the continuum model considering both the lattice relaxation effects and the non-local moir\'e potential terms. (d) Schematic to construct a generic continuum model. We distinguish the twist angle independent intrinsic properties and the twist angle dependent external inputs. The intrinsic properties can be fitted via small-size DFT calculation and the external inputs are obtained by Deep Potential Molecule Dynamics (DPMD). Based on one set of universal parameters, the electronic properties of the given moir\'e superlattice can be accurately obtained. Subfigure (a-c) are extracted from Bo Xie and Jianpeng Liu, Phys. Rev. B, Vol.108, 094115, 2023 \cite{xie_phonon_prb23}; licensed under a Creative Commons Attribution (CC BY) license.}
\label{fig:structure_band}
\end{figure*}

\subsection{Continuum model for twisted TMD}
While the BM model was tailored to graphene's linear Dirac dispersion, TMDs exhibit fundamentally different electronic structures: parabolic valence and conduction bands near the band edges and strong spin-orbit coupling (SOC), which induces opposite spin splittings for $K$ and $K'$ valleys near the valence band maxima. These differences demanded the development of TMD-specific continuum models. In 2019, Wu et al. proposed the first version continuum model for twist MoTe$_{2}$ \cite{wu-TI-twisted-tmd-prl19}. Focusing on the valence band maximum, they consider the leading order Fourier components of the intralayer and interlayer moir\'e potential. The Hamiltonian is given as:
\begin{align}
\scalebox{0.85}{
$\begin{aligned}
\begin{split}
H_{\textbf{K}}^{tMoTe_{2}}=\left[\begin{array}{cc}
				-\frac{\hbar^{2}(\textbf{k}-\textbf{K}^{(1)})^{2}}{2m^{*}}+\Delta_{(1)}(\textbf{r}) & \Delta_{T}(\textbf{r}) \\
				 \Delta_{T}^{\dagger}(\textbf{r}) & -\frac{\hbar^{2}(\textbf{k}-\textbf{K}^{(2)})^{2}}{2m^{*}}+\Delta_{(2)}(\textbf{r})\cdot\sigma^{\mu}
\end{array}\right],
\end{split}
\end{aligned}$%
}
\end{align}
where $\Delta_{(l)}(\textbf{r})\!=\!2V\sum_{j} \cos(\textbf{G}_{j}^{M}\cdot\textbf{r}+l\psi)$ represents the intralayer moir\'e potential, where $V$ describe the amplitude. $\Delta_{T}(\textbf{r})\!=\!w(1+e^{-i\textbf{G}_{1}^{M}\cdot\textbf{r}}+e^{-i(\textbf{G}_{1}^{M}+\textbf{G}_{2}^{M})\cdot\textbf{r}})$ represents the interlayer moir\'e potential, with $w$ as its amplitude. $m^{*}$ represents the effective mass of the band edge. The model coefficients $\{V, w, \psi\}$ are fitted from DFT calculations of lateral shifted bilayer MoTe$_{2}$ system without twist. Subsequent extensions incorporate the higher Fourier components of moir\'e potentials, which turn out to be necessary to describe more details of the moir\'e potentials \cite{fuliang_charge_order_ttmd_prb21, fuliang_ah_fci_ttmd_prb23, zhangyang_transfer_learning_continuum_tmote2_cp24,xiao-fqah-arxiv23, zhangyang_multiple_chern_band_tmote2_prl25, wuquansheng_dft_tmote2_prb24}. These extensions rely on large-scale DFT calculations of twisted MoTe$_{2}$ superlattices (with thousands of atoms) to fit higher-order potential amplitudes. Despite these successes, TMD continuum models suffer from three critical limitations that remain an active research area. Firstly, one approach assumes a single set of parameters can describe low energy properties across all twist angles, implying the moir\'e potential is invariant to twist. However, this is physically unrealistic: twist angle modifies the moir\'e unit-cell size and interlayer overlap, which should alter potential amplitudes. Others prefer to fit model parameters separately for each twist angle, but resulting values vary non-monotonically. This non-monotonicity casts doubt on the physical interpretability of the parameters, as fundamental quantities like potential amplitude should depend smoothly on twist angle. Secondly, some researchers improve the accuracy of continuum models by incorporating large-scale DFT calculations, which dramatically increases computational cost. Thirdly, despite reproducing the band structure, there is a need to establish a clear linkage between the parameters in continuum models and the underlying microscopic physics. Additional validation of these parameters is also essential—for example, the parameters should vary smoothly with twist angle and exhibit a smooth profile in both real and reciprocal space. A more recent approach constructs the continuum model by numerically projecting the DFT Hamiltonian in numerical atom orbitals onto the continuum model basis \cite{zhang-wu-continuum-arxiv24}. This method achieves unprecedented agreement with DFT band structures, as it directly incorporates atomistic details into the continuum framework. However, it requires performing large-scale DFT calculations for every twist angle of interest.

\subsection{Generic continuum model}
There is demand for a generic continuum model for arbitrary moir\'e superlattice with clear physical meanings. A desirable continuum model construction would distinguish between electrons' intrinsic properties and the system's external inputs \cite{xie-continuum_model-arxiv-2025,xiao-continuum-arxiv25}, as schematically illustrated in Fig.~\ref{fig:structure_band}(d). From our perspective, the intrinsic properties include electrons' the microscopic kinetic energy and the atomic lattice potential exerted to the electrons; these properties depend on the constituent materials' symmetries, atomic species, and atomic orbitals, but remain invariant across different moir\'e system sizes (twist angle $\theta$). By contrast, the external input variable is $\theta$-dependent the lattice relaxation patterns, coupling to both the kinetic energy and moir\'e potential terms via $\theta$-invariant coefficients. 

In such a context, we try to develop a generic continuum model for low-energy electronic properties of moir\'e superlattice systems \cite{xie-continuum_model-arxiv-2025}.  We focus on systems where low-energy physics is dominated by valence band maximum (VBM) or the conduction band minimum (CBM) at $\textbf{K}$ point in the reciprocal space of some hexagonal unit-cell, common in moir\'e graphene and TMDs. It is straightforward to expand around any point (``valley'') in reciprocal space. The following discussions mostly follow those reported in our own work \cite{xie-continuum_model-arxiv-2025}. Similar formalism is also reported in Ref.~\cite{xiao-continuum-arxiv25}.

Neglecting intervalley scattering, the non-interacting Hamiltonian takes block-diagonal form in valley space. For $\textbf{K}$ valley, the effective continuum model can be expressed as:
\begin{align}
\begin{split}
H^{\textbf{K}}=\left[\begin{array}{cc}
				T^{(1)}+V^{(1)} & W \\
				W^{\dagger} & T^{(2)}+V^{(2)}
\end{array}\right].
\end{split}
\end{align}
$T^{(l)}$ and $V^{(l)}$ are the kinetic energy and the intralayer moir\'e potential from the $l$-th layer; while $W$ is the interlayer moir\'e potential. Each term expands as power series in wavevector $\mathbf{k}$ and distortion fields $\{\mathbf{u}^{\pm}, h^{\pm}\}$. 
\begin{align}
&\mathcal{F}_{\lambda,\lambda'}(\textbf{k})=\sum_{\alpha}^{\infty}\sum_{\beta}^{\alpha}\bar{\mathcal{F}}_{\lambda,\lambda'}^{\alpha,\beta} (\{\mathbf{u}^{\pm},h^{\pm}\})\,k_{x}^{\beta}k_{y}^{\alpha-\beta},
\end{align}
where $\mathcal{F}\!=\!\{T^{(l)}, V^{(l)}, W\}$, and $\mathbf{k}=\tilde{\mathbf{k}}+\mathbf{G}+\mathbf{K}^{(l)}-\mathbf{K}$ is the wavevector expanded around the untwisted $\mathbf{K}$ point, $\tilde{\mathbf{k}}$ is the wavevector within the first moir\'e Brillouin zone, $\textbf{G}$ is the reciprocal vector. Integer $\alpha$ is the total order of the Taylor expansion, integer $\beta$ is the order of $k_{x}$. $\bar{\mathcal{F}}^{\alpha,\beta}_{\lambda,\lambda'}$ can be further expanded as series in strain fields:
\begin{align}
\bar{\mathcal{F}}_{\lambda,\lambda'}^{\alpha,\beta}=&\bar{\mathcal{F}}_{\lambda,\lambda'}^{\alpha,\beta,0}+\sum_{\{S\}}\sum_{\textbf{G}}\left[\frac{\partial}{\partial S_{\textbf{G}}}\bar{\mathcal{F}}_{\lambda,\lambda'}^{\alpha,\beta}\right]S_{\textbf{G}}\nn
&+\sum_{\{S^{(1)}\},\{S^{(2)}\}}\sum_{\textbf{G}_{1},\textbf{G}_{2}}\left[\frac{\partial^{2}}{\partial S^{(1)}_{\textbf{G}_{1}}\partial S^{(2)}_{\textbf{G}_{2}}}\bar{\mathcal{F}}_{\lambda,\lambda'}^{\alpha,\beta}\right]S^{(1)}_{\textbf{G}_{1}}S^{(2)}_{\textbf{G}_{2}}
\end{align}
where $\{S\}=\{\textbf{u}^{\pm}, h^{\pm}\}$ represents different types of distortion fields and $S_{\G}$ are their corresponding Fourier components.

We distinguish moir\'e potential into momentum-independent (local) and momentum-dependent (non-local) components. Real-space non-locality means Fourier components of the potential depend not only on the moir\'e reciprocal vector $\mathbf{Q}_j$, but also on Bloch state incident wavevector $\mathbf{k}$. Here, we expand the nonlocal moir\'e potential in polar coordinate of $\mathbf{k}$, which takes the form:
\begin{align}
V_{\mathbf{Q}_j}(\textbf{k})=\sum_{n,m} V^{m,n}_{\mathbf{Q}_j}\,\mathcal{T}_{n}(|\textbf{k}|L_{s}\eta) e^{i m\phi},
\end{align}
where $\mathcal{T}_{n}$ is the $n$-th Chebyshev polynomial. The factor $\eta$ sets the cutoff for the spatial extent of the non-local moir\'e potential to be of order $\eta L_s$. This is technically necessary because the expansion in Chebyshev polynomials is complete only on a finite interval.

\subsubsection{Correction to kinetic energy induced by strain} 
We analyze the strain induced correction to kinetic energy by linking lattice distortions to kinetic energy changes via tight-binding theory. Consider a unit-cell containing $N_{W}$ Wannier functions. We define the displacement vector from $n$-th Wannier function in the ``home'' cell to the neighboring $m$-th Wannier function in the $j$th unit-cell $\textbf{r}_{nm}^{j}$. A three-dimensional lattice strain tensor $\hat{\mathbf{u}}$ modifies displacement vectors $\textbf{r}_{nm}^{j}$ to $(\textbf{r}_{nm}^{j})'\!=\!(1+\hat{\mathbf{u}})\textbf{r}_{nm}^j$ and leads to a shift in the hopping amplitude $\delta t^{n,m}(\textbf{r})\!=\!t^{n,m}(\textbf{r}')-t^{n,m}(\textbf{r})$. The tight binding Hamiltonian is given by: $t^{n,m}(\textbf{k})\!=\!\sum_{j}t^{n,m}e^{i\,\textbf{k}\cdot\textbf{r}_{nm}^j}$. We expand the tight binding model around the band edge wavevector:
\begin{align}
\scalebox{0.9}{
$\begin{aligned}
&t^{n,m}(\textbf{k}+\textbf{K})=\sum_{j}(t^{n,m}(\textbf{r}_{nm}^j)+\delta t^{n,m}(\textbf{r}_{nm}^j))e^{i(\textbf{K}+\textbf{k})\cdot\textbf{r}_{nm}^j}\\
\approx&\sum_{j}\,e^{i\textbf{K}\cdot\textbf{r}_{nm}^j}\Big(\,t^{n,m}(\textbf{r}_{nm}^j)+\delta t^{n,m}(\textbf{r}_{nm}^j)(1+i\textbf{k}\cdot\textbf{r}_{nm}^j+\dots)\,\Big).
\label{eq:strainkinetic}
\end{aligned}$%
}
\end{align}
The first term is the unstrained band-edge Hamiltonian; subsequent terms are strain-induced corrections, with the expansion in $\textbf{k}$ capturing the low-energy dispersion. System's symmetries further constrain expansion parameters.

\subsubsection{Correction to intralayer moir\'e potential induced by strain}
For local intralayer moir\'e potentials, the twisted adjacent layer induces a position-dependent potential $V^{(l)}(\bm{\delta}(\textbf{r}))$ on the $l$-th layer, where $\bm{\delta}(r)\!=\!\bm{\delta}_{0}+\textbf{u}^{-}(\textbf{r})+(h^{-}(\textbf{r}))\textbf{e}_{z}$ is the local displacement. Lattice relaxation alters interlayer atomic positions and thus the potential can be formally expressed as $V^{(l)}(\bm{\delta}(\textbf{r}))\!=\!V^{(l)}(\bm{\delta}_{\parallel}(\textbf{r}),h^{-}(\textbf{r}))$, where $\bm{\delta}_{\parallel}(\textbf{r})=\bm{\delta}_{0}(\textbf{r})+\textbf{u}^{-}(\textbf{r})$. Fourier expanding $V^{(l)}(\bm{\delta}_{\parallel}(\textbf{r}),h^{-}(\textbf{r}))$ in terms of the in-plane distortion gives:
\allowdisplaybreaks
\begin{align}
\scalebox{0.9}{
$\begin{aligned}
&V^{(l)}(\bm{\delta}_{\parallel}(\textbf{r}),h^{-}(\textbf{r}))\\
=&\sum_{j}V^{(l)}_{\textbf{G}_j}(h^{-}(\textbf{r}))e^{i\textbf{a}_{j}^{*}\cdot\bm{\delta}_{\parallel}(\textbf{r})}=\sum_{j}V^{(l)}_{\textbf{G}_j}(h^{-}(\textbf{r}))e^{i\textbf{a}_{j}^{*}\cdot(\bm{\delta}_{0}(\textbf{r})+\textbf{u}^{-}(\textbf{r}))}\\
=&\sum_{j}V^{(l)}_{\textbf{G}_j}(h^{-}(\textbf{r}))e^{i\textbf{G}_{j}\cdot\textbf{r}}\prod_{\textbf{G}^{u}_{m}}\sum_{n_{m}}\frac{1}{n_{m}!}(i\textbf{a}_{j}^{*}\cdot\textbf{u}^{-}_{\textbf{G}^{u}_{m}})^{n_{m}}e^{i n_{m}\textbf{G}^{u}_{m}\cdot\textbf{r}}\\
=&\sum_{j}\sum_{n_{m_{1}},\dots}V^{(l)}_{\textbf{G}_j}(h^{-}(\textbf{r}))\frac{\left(i\textbf{a}^{*}_{j}\cdot\textbf{u}^{-}_{\textbf{G}^{u}_{m_{1}}}\right)^{n_{m_{1}}}}{n_{m_{1}}!}\frac{\left(i\textbf{a}^{*}_{j}\cdot\textbf{u}^{-}_{\textbf{G}^{u}_{m_{2}}}\right)^{n_{m_{2}}}}{n_{m_{2}}!}\dots\\
&\ \ \ \ \ \dots e^{i\textbf{G}_{j}\cdot\textbf{r}}e^{i(n_{m_1}\textbf{G}^{u}_{m_1}+n_{m_2}\textbf{G}^{u}_{m_2}+\dots)\cdot\textbf{r}}.
\end{aligned}$%
}
\end{align}
The product over $\textbf{G}^{u}_{m}$ and sum over $n_{m}$ account of the contribution of multiple strain harmonics, with $n_{m}$ account for the order of expansion in the strain fields. Setting $n_{m_{i}}=0$ recovers the unrelaxed moir\'e potential.  Out-of-plane distortion $h^{-}(\mathbf{r})$ weakly affects intralayer potential (vs. interlayer). We neglect its spatial variation, taking $h^{-}(\textbf{r})\rightarrow d_{0}$. Here $d_{0}$ is the average interlayer distance. When the average layer spacing deviates from $d_{0}$ by a small amount $\delta d$, say, due to the change of twist angle, the intralyer potential undergos a corresponding change:
\begin{align}
V^{(l)}_{\textbf{G}}(d_{0}+\delta d)\approx& V^{(l)}_{\textbf{G}}(d_{0})+\left.\frac{\partial V^{(l)}_{\textbf{G}}}{\partial d'}\right|_{d'=d_{0}}\delta d +\dots\nn
=&V^{(l)}_{\textbf{G}}(d_{0})+ V^{(l),1}_{\textbf{G}}(d_{0})\delta d+\dots.
\label{eq:vd}
\end{align}
Here, $V^{(l),1}_{\textbf{G}}(d_{0})\!=\!\partial V^{(l)}_{\textbf{G}}/\partial \left. d'\right|_{d'=d_{0}}$ quantifies the linear dependence of the intralayer moir\'e potential to the average interlayer distance.

\subsubsection{Correction to interlayer moir\'e potential induced by strain} 
Again, we first focus on the local interlayer moir\'e potential, deriving its explicit form via microscopic interlayer hopping. We begin by considering the microscopic hopping process between atomic sites in adjacent layers. Consider atomic-site hopping between adjacent layers, with amplitude $t(\textbf{r}+d\textbf{e}_{z})$, under two-particle approximation. We define its 2D Fourier transformation as
$W(\textbf{q})=\int \mathrm{d}^{3}r\, t(\textbf{r}+d\textbf{e}_{z})e^{-i\textbf{q}\cdot\textbf{r}}/S_{0}d_{0}$,
where $S_{0}$ is the unit-cell area and $d_{0}$ is the average interlayer distance. The interlayer hopping term is expressed as:
\begin{align}
\scalebox{0.85}{
$\begin{aligned}
&\bra{\textbf{k}',X',l'}U\ket{\textbf{k},X,l}=\sum_{\textbf{g}^{(1)},\textbf{g}^{(2)}}\sum_{n_{1},\dots}\sum_{n_{h,1},\dots}\sum_{n_{1}'\dots}\sum_{n'_{h,1}\dots}e^{-i(\textbf{g}^{(1)}\cdot\bm{\tau}_{X}-\textbf{g}^{(2)}\cdot\bm{\tau}_{X'})}\\
&\ \ \ \ \ \ \ \ \ \gamma(\textbf{Q})\delta_{\textbf{k}+\textbf{g}^{(1)}+n_{1}\textbf{q}_{1}+n_{h,1}\textbf{q}_{h,1}+\dots,\textbf{k}'+\textbf{g}^{(2)}-n_{1}\textbf{q}_{1}-n_{h,1}\textbf{q}_{h,1}+\dots},
\end{aligned}$%
}
\end{align}
where $\textbf{Q}=\textbf{Q}_{\parallel}+p_{z}\textbf{e}_{z}$, and the in-plane wavevector
$\textbf{Q}_{\parallel}=\textbf{k}+\textbf{g}^{(1)}+n_{1}\textbf{G}^{u}_{1}+n_{h,1}\textbf{G}^{h}_{1}+n_{2}\textbf{G}^{u}_{2}+n_{h,2}\textbf{G}^{h}_{2}+\dots
=\textbf{k}'+\textbf{g}^{(2)}-n'_{1}\textbf{G}^{u}_{1}-n'_{h,1}\textbf{G}^{h}_{1}-n'_{2}\textbf{G}^{u}_{2}-n'_{h,2}\textbf{G}^{h}_{2}+\dots$, $\{\textbf{G}^{u}_{1},\textbf{G}^{u}_{2},\textbf{G}^{h}_{1},\textbf{G}^{h}_{2},\dots\}$ are moir\'e reciprocal vectors. The effective interlayer hopping amplitude is expressed as:
\begin{align}
&\gamma(\textbf{Q}_{\parallel}+p_{z}\textbf{e}_{z})\nn
\approx&\frac{d_{0}}{2\pi}\int\mathrm{d}p_{z}\,W(\textbf{Q}_{\parallel}+p_{z}\textbf{e}_{z})e^{i\,p_{z}d_{0}}\nn
&\times\frac{\left[i\textbf{Q}_{\parallel}\cdot\textbf{u}^{-}_{\textbf{G}^{u}_{1}}\right]^{n_{1}+n'_{1}}}{n_{1}!n'_{1}!}\frac{\left[i\textbf{Q}_{\parallel}\cdot\textbf{u}^{-}_{\textbf{G}^{u}_{2}}\right]^{n_{2}+n'_{2}}}{n_{2}!n'_{2}!}\dots\nn
&\times\frac{\left[i\,p_{z}h^{-}_{\textbf{G}^{h}_{1}}\right]^{n_{h,1}+n'_{h,1}}}{n_{h,1}!n'_{h,1}!}\frac{\left[i\,p_{z}h^{-}_{\textbf{G}^{h}_{2}}\right]^{n_{h,2}+n'_{h,2}}}{n_{h,2}!n'_{h,2}!}\dots,
\end{align}
where $n_{m}$, $n_{m}'$, $n_{h,m}$ and $n_{h,m}'$ are the expansion orders. 
The integration over $p_{z}$ can be carried out using the trick of integration by part. We provide the zeroth, first and second order:
\begin{align}
\scalebox{0.8}{
$\begin{aligned}
&\frac{d_{0}}{2\pi}\int \mathrm{d} p_{z}e^{i p_{z}d_{0}}W(\textbf{Q}_{j}+p_{z}\textbf{e}_{z})\\
=&\frac{1}{S_{0}}\int \mathrm{d}^{2}r\,t(\textbf{r}+d_{0}\textbf{e}_{z})e^{-i\textbf{Q}_{j}\cdot\textbf{r}}=\tilde{W}_{0}\\
&\frac{d_{0}}{2\pi}\int\mathrm{d}p_{z}\,W(\textbf{Q}_{\parallel}+p_{z}\textbf{e}_{z})e^{i\,p_{z}d_{0}}\times\left[i\,p_{z}\right]\\
=&\frac{1}{S_{0}}\int\mathrm{d}^{2}r\left.\frac{\mathrm{d}}{\mathrm{d}d'}t(\textbf{r}+d'\textbf{e}_{z})\right|_{d'=d_{0}}e^{-i\textbf{Q}\cdot\textbf{r}}=\tilde{W}_{1}\\
&-\frac{d_{0}}{2\pi}\int \mathrm{d} p_{z}e^{i p_{z}d_{0}}W(\textbf{Q}_{j}+p_{z}\textbf{e}_{z})\times i p_{z}h^{-}_{\textbf{G}^{h}_{m_1}}i p_{z}h^{-}_{\textbf{G}^{h}_{m_2}}\\
=&-\frac{1}{S_{0}}\int \mathrm{d}^{2}r \left.\frac{\mathrm{d}^{2}}{\mathrm{d}^{2} d'}t(\textbf{r}+d'\textbf{e}_{z})\right|_{d'=d_{0}}e^{-i\textbf{Q}_{j}\cdot\textbf{r}}h^{-}_{\textbf{G}^{h}_{m_1}}h^{-}_{\textbf{G}^{h}_{m_2}}=\tilde{W}_{2}\;.
\end{aligned}$%
}
\end{align}
Combining these contributions, the interlayer hopping amplitude is expressed as:
\begin{align}
\scalebox{0.8}{
$\begin{aligned}
&\gamma(\textbf{Q}_{\parallel}+p_{z}\textbf{e}_{z})\\
=&\tilde{W}_{0}(d_{0})\frac{\left[i\textbf{Q}_{\parallel}\cdot\textbf{u}^{-}_{\textbf{G}_{m_1}}\right]^{n_{m_1}+n'_{m_1}}}{n_{m_1}!n'_{m_1}!}\frac{\left[i\textbf{Q}_{\parallel}\cdot\textbf{u}^{-}_{\textbf{G}_{m_2}}\right]^{n_{m_2}+n'_{m_2}}}{n_{m_2}!n'_{m_2}!}\dots\\
&+\tilde{W}_{1}(d_{0})h_{\textbf{G}^{h}_{1}}^{-}\frac{\left[i\textbf{Q}_{\parallel}\cdot\textbf{u}^{-}_{\textbf{G}_{m_1}}\right]^{n_{m_1}+n'_{m_1}}}{n_{m_1}!n'_{m_1}!}\frac{\left[i\textbf{Q}_{\parallel}\cdot\textbf{u}^{-}_{\textbf{G}_{m_2}}\right]^{n_{m_2}+n'_{m_2}}}{n_{m_2}!n'_{m_2}!}\dots\\
&+\tilde{W}_{2}(d_{0})h_{\textbf{G}^{h}_{1}}^{-}h_{\textbf{G}^{h}_{2}}^{-}\frac{\left[i\textbf{Q}_{\parallel}\cdot\textbf{u}^{-}_{\textbf{G}_{m_1}}\right]^{n_{m_1}+n'_{m_1}}}{n_{m_1}!n'_{m_1}!}\frac{\left[i\textbf{Q}_{\parallel}\cdot\textbf{u}^{-}_{\textbf{G}_{m_2}}\right]^{n_{m_2}+n'_{m_2}}}{n_{m_2}!n'_{m_2}!}\dots.
\end{aligned}$%
}
\end{align}
This expansion explicitly realizes the strain-dependent interlayer moir\'e potential, with each term quantifying how lattice distortions, both in-plane and out-of-plane, modify the hopping amplitude. 
Notably, the amplitude of $\tilde{W}_{0}$, $\tilde{W}_{1}$ and $\tilde{W}_{2}$ are fixed for given $d_{0}$, as they depend only on the microscopic hopping $t$ and the average interlayer distance. A small shift $\delta d$ from a reference $d_{0}$ modifies $\tilde{W}_{0}$ through a Taylor expansion:
\begin{align}
&\tilde{W}_{0}(d_{0}+\delta d)\nn
\approx&\tilde{W}_{0}(d_{0})+\left.\frac{\partial\tilde{W}}{\partial d'}\right|_{d'=d_{0}}\delta d+\frac{1}{2}\left.\frac{\partial^{2}\tilde{W}}{\partial d'^{2}}\right|_{d'=d_{0}}\delta d^{2}+\dots\nn
=&\tilde{W}_{0}(d_{0})+\tilde{W}_{1}(d_{0})\delta d+\frac{1}{2}\tilde{W}_{2}(d_{0})\delta d^{2}+\dots.
\label{eq:td}
\end{align}
This expansion reveals a dual role for $\tilde{W}_{1}$ and $\tilde{W}_{2}$: they not only describe the correction of corrugation effects to the interlayer moir\'e potential, but also describe the variation of interlayer hopping parameters at different $d_{0}$. 

The average interlayer distance $d_{0}$ is not a fixed value, but is determined by the lattice relaxations of the moir\'e superlattice system.  As the system size (twist angle) varies, the local stacking pattern is modified, altering the average interlayer distance $d_{0}$. This, in turn, builds up the twist angle dependent moir\'e potential.

After formulating the effective continuum model, we derive a semi-analytical continuum model characterized by a set of parameters: $\mathcal{P}\!=\!$ $\{V^{(l)}_{\textbf{G}}$, $V^{(l),1}_{\textbf{G}}$, $\tilde{W}_{0}$, $\tilde{W}_{1}$, $\tilde{W}_{2},\dots\}$. As shown in Eq.~\eqref{eq:vd} and Eq.~\eqref{eq:td}, these parameters encode the dependence of moir\'e potentials on the average interlayer distance $d_{0}$, enabling to describe systems with different size (e.g., varying twist angle) with a single consistent parameter set. The parameters can be fitting from large-twist-angle ($\theta$) DFT calculation, while the lattice distortion are solved using Deep Learning Molecule Dynamics. This approach enables accurate descriptions of low-energy physics without relying on computationally expensive large-scale DFT simulations.

\subsection{Applications}
To demonstrate the power of this method, we have applied it to twisted MoTe$_2$ systems in our recent work \cite{xie-continuum_model-arxiv-2025}, where we present systematic evaluations of band structures, real-space charge distributions, and Chern numbers across multiple twist angles. Importantly, the results from our generic continuum model show strong consistency with the corresponding DFT calculations for these angles. Moreover, a recent scanning tunneling spectroscopy study \cite{liu-tmote2STM-arxiv-2025} was conducted on several twisted MoTe$_2$ devices at different twist angles, in which well-defined peaks are observed at the MX/MM points, in agreement with the results calculated by the generic continuum model.

\section{Hartree-Fock mean-field approximations}
Once a continuum model is obtained, Hartree-Fock (HF) approximations are the simplest approach to simplify $e$-$e$ Coulomb interactions by factorizing the quartic four-operator form into a quadratic form, known as HF potential, using the mean value of density matrix elements and neglecting the fluctuations with respect to the mean value of density matrix elements. The resulting quadratic HF Hamiltonian—composed of the non-interacting Hamiltonian plus HF potentials—yields wavefunctions of the Slater determinant type, similar to those of the non-interacting case, which can then be interpreted in a physically transparent way. To achieve self-consistency, the mean values of density matrix elements calculated from the eigenstates of the HF Hamiltonian (the HF states) must coincide with those used to construct the HF potential. This method has been widely applied to moir\'e systems and has proven remarkably successful in explaining a range of experimental observations, particularly the cascade transitions at different fillings observed in TBG \cite{Yazdani-cascade-nature20,Ilani-cascade-nature20,kang-cascade-tbg-prl21}. The physics of correlated states for integer filled moir\'e flat bands is reminiscent of the problem of quantum Hall ferromagnetism in the strong coupling limit where exchange interactions dominate over all the other $e$-$e$ interactions \cite{ledwith-strongcoupling-tbg,khalaf-soft-arxiv20,khalaf-skyrmionTBG-sciadv-2021}. The accuracy of HF calculations has also been supported by density matrix renormalization group method \cite{zaletel-dmrg-prb20,bultinck-tbg-strain-prl21,kang-tbg-dmrg-prb20}, many-body perturbation theory \cite{lu-MBPTmoire-arxiv-2025} and exact diagonalization calculations \cite{regnault-tbg-ed,macdonald-tbg-ed-prl21,yu-R5GED-prb-2025}. The last two methods will be discussed in the next two sections.

In this section, we introduce the general formalism of unconstrained Hartree-Fock approximations to $e$-$e$ interactions in moir\'e systems. Then, we show how to reduce computational costs by projecting the calculations into an active low-energy subspace within which self-consistent loops are systematically performed. Several techniques are presented to account for the influence of remote bands outside the active window even after projection. We also discuss different conventions of normal ordering used in the literature for HF calculations, which can yield qualitatively distinct HF phase diagrams. These differences can often be traced back to how remote band effects are treated. Comparisons with experiments suggest that a proper treatment of these remote bands is essential. To address this issue decisively, we propose performing HF calculations with all the bands included in the continuum model, an approach we call all-band HF. Finally, we discuss the limitations of HF approximations, which motivate the use of more advanced methods such as many-body perturbation theory.

\subsection{General formalism}
\label{sec:HF}
Before introducing any mathematical formalism, let us lay down an assumption that allows to treat the $e$-$e$ Coulomb interactions within continuum model. Specifically, we regard the valence electrons lying outside the plane-wave cutoff of the continuum model as spectators, analogous to core electrons. This assumption is justified because the continuum model provides a reliable description of the non-interacting band structure in the low-energy regime (on the order of tens of meV), whereas the plane-wave cutoff energy typically exceeds 1\;eV. We further assume the presence of a homogeneous, frozen positive background charge to guarantee charge neutrality of the system. Under these conditions, we can begin directly with the normal-ordered form of the Coulomb interaction. A detailed derivation of the formulas presented below can be found in the literature and the corresponding supplementary materials \cite{zhang-liu-tbg-prl22,zhang-tbmg-prl22,guo-HFfci-prb-2024}.

Let us first write the $e$-$e$ Coulomb interactions in its general second quantized form
\begin{equation}
\hat{V}=\frac{1}{2}\int d^2 r  d^2 r' \sum _{\sigma, \sigma '} \hat{\psi}_\sigma ^{\dagger}(\br)\hat{\psi}_{\sigma '}^{\dagger}(\br ') V_c (|\br -\br '|) \hat{\psi}_{\sigma '}(\br ') \hat{\psi}_{\sigma}(\br)
\label{eq:coulomb_r}
\end{equation}
where $\hat{\psi}^{(\dagger)}_{\sigma}(\br)$ is real-space electron annihilation (creation) operator at $\br$ with spin $\sigma$ and the Coulomb interaction $V_c$ is assumed to be rotationally invariant. This can be further expressed in the atomic basis 
\begin{equation}
\begin{split}
    \hat{V}=&\frac{1}{2}\sum _{i i' j j'}\sum_{l l' m m'}\sum _{\alpha \alpha '\beta \beta ' }\sum _{\sigma \sigma '}  V^{\alpha \beta l m \sigma , \alpha ' \beta ' l' m' \sigma '} _{ij,i'j'} \\
    &\times \hat{c}^{\dagger}_{i, \sigma l \alpha}\hat{c}^{\dagger}_{i', \sigma ' l' \alpha '} \hat{c}_{j', \sigma ' m' \beta '} \hat{c}_{j, \sigma m \beta}\;, 
\end{split}
\label{eq:coulomb_alpha}
\end{equation}
where the operator $\hat{c}_{i, \sigma l \alpha}^{(\dagger)}$ annihilates (creates) an electron of spin $\sigma$ at layer $l$ on sublattice $\alpha$ in the atomic unit-cell $i$ with a family of localized atomic Wannier functions  \{$\phi_{l,\alpha}(\br)\}$. The form factor for Coulomb interactions reads
\begin{align}
& V^{\alpha \beta l m \sigma , \alpha ' \beta ' l' m' \sigma '} _{ij,i'j'} = \int d^2 r d^2 r'  V_c (|\br -\br '|)  \nonumber \\
&\quad\times \phi^*_{l,\alpha} (\mathbf{r}-\mathbf{R}_i-\bm{\tau}_{l,\alpha}) \phi_{m,\beta} (\mathbf{r}-\mathbf{R}_j- \bm{\tau}_{m,\beta})  \nonumber \\
&\quad\times \phi^*_{l', \alpha  '}(\br'-\mathbf{R}_i'-\bm{\tau}_{l' , \alpha '}) \phi_{m', \beta  '}(\br'-\mathbf{R}_j'-\bm{\tau} _{m', \beta '}) .
\end{align}
where the relative position $\bm{\tau}_{l,\alpha}$ is the center of the Wannier functions within the $i$-th atomic unit-cell labeled by lattice vector $\mathbf{R}_i$.

Thanks to the localization of the wavefunction $\phi_{l,\alpha}(\br)$, we further assume that the ``density-density'' like interaction is dominant in the system, i.e., the form factor in the atomic basis 
\begin{equation}
    V^{\alpha \beta l m \sigma , \alpha ' \beta ' l' m' \sigma '} _{ij,i'j'}\approx V^{\alpha \alpha l l \sigma , \alpha ' \alpha ' l' l' \sigma '} _{ii,i'i'}\equiv V_{i \sigma l \alpha  ,i' \sigma ' l' \alpha '} \; ,
\end{equation}
then the Coulomb interaction is simplified to
\begin{equation}
    \begin{split}
        \hat{V}=&\frac{1}{2}\sum _{i l \alpha \neq i' l' \alpha '}\sum _{\sigma \sigma '} V_{i l \alpha,i' l' \alpha '} \hat{c}^{\dagger}_{i, \sigma l \alpha} \hat{c}^{\dagger}_{i', \sigma ' l' \alpha '} \hat{c}_{i', \sigma ' l' \alpha '}\hat{c}_{i, \sigma l \alpha} \\
        &+ U_0 \sum _{i l \alpha } \hat{c}^{\dagger}_{i, \uparrow l \alpha} \hat{c}^{\dagger}_{i, \downarrow  l \alpha } \hat{c}_{i, \downarrow   l \alpha }\hat{c}_{i, \uparrow l \alpha} \; ,
    \end{split}
\end{equation}
where we split Coulomb interactions into an intersite and an on-site Hubbard-like interaction with amplitude $U_0$. 

To model effectively the long-wavelength screening effects to the $e$-$e$ Coulomb interactions, we can use the Thomas-Fermi screening form of the intralayer Coulomb interactions, whose Fourier transform is expressed as 
\begin{equation}
    V (\mathbf{q})=\frac{e^2}{2 \Omega_0 \epsilon_r \epsilon_0 \sqrt{q^2+\kappa^2}}
 \label{eq:V_thomasfermi}
\end{equation}
where $\Omega _0 = \sqrt{3}L_s^2 /2 $ is the area of the triangular moir\'e superlattice's primitive cell with moir\'e lattice constant $L_s$, $\epsilon_0$ the vacuum permittivity, $\epsilon_r$ the homogeneous static dielectric constant and $\kappa$ inverse screening length. A double-gate screening form is also widely used in literature
\begin{equation}
    V(\mathbf{q})=\frac{e^2\tanh(|\mathbf{q}|d_s)}{\,2\Omega _0 \epsilon_r \epsilon _0 |\mathbf{q}|} \; ,
    \label{eq:V_doublegate}
    \end{equation}
where the screening length $d_s$ can be considered as the thickness of the device. For the Coulomb interactions between electrons from different layers, we use
\begin{equation}
    V_{ll'}(\mathbf{q})=\frac{e^2}{2 \Omega_0 \epsilon_r \epsilon_0 q} e^{-q|l-l'|d_0}
    \label{eq:V_interlayer}
\end{equation}
with $l \neq l'$ and $d_0$ the average distance between two adjacent layers. The divergence at $q=0$ should not worry us since it is physically regularized by the compensation with positive charge background. This allows us to exclude the point $q=0$ in the calculations.

Calculations show that adopting different functional forms of the Coulomb interaction or slightly varying the screening length does not lead to qualitatively different HF results, provided the dielectric constant is kept fixed. This brings us to the central role of the dielectric constant. The static dielectric constant, $\epsilon_r$, has two contributions: environmental screening and intrinsic screening. The environmental contribution—always greater than unity—arises from the substrate of the device and from high-energy spectator electrons, such as core electrons and those lying beyond the continuum model cutoff. The intrinsic contribution originates from the homogeneous static screening among the low-energy electrons described within the continuum model.

In practice, $\epsilon_r$ is usually treated as a fitting parameter, since determining it \textit{ab initio} in moir\'e systems is extremely challenging. Its value also depends on the number of layers in the system. For instance, in hBN-encapsulated twisted graphene systems, it is common to set $\epsilon_r = \epsilon_{\rm hBN}$ for monolayers, while for twisted multilayer systems $\epsilon_r$ may vary from 4 to 20.

Back to Eq.~\eqref{eq:coulomb_alpha}, the inter-site Coulomb interactions can be divided into the intra-valley term and the inter-valley term. Take twisted graphene as an example. The intra-valley term $\hat{V}^{\text{intra}}$ can be expressed in the Fourier basis as
\begin{equation}
    \begin{split}
\hat{V}^{\rm{intra}}=&\frac{1}{2N_s}\sum_{\alpha\alpha ', l l'}\sum_{\mu\mu ',\sigma\sigma '}\sum_{\bk \bk ' \q} V_{l l'}(\q) \hat{c}^{\dagger}_{\sigma \mu l \alpha}(\bk+\q)  \\
&\times \hat{c}^{\dagger}_{\sigma' \mu ' l' \alpha '}(\bk ' - \q) \hat{c}_{\sigma ' \mu ' l' \alpha '}(\bk ')\hat{c}_{\sigma \mu l \alpha}(\bk)\;,
    \end{split}
\label{eq:V-intra}
\end{equation}
where $N_s$ is the total number of the superlattice's sites and the momentum transfer occurs within the same valley for $q \ll |\mathbf{K}-\mathbf{K}'| $. The operator $\hat{c}^{(\dagger)}_{\sigma \mu l \alpha}(\bk)$ annihilates (creates) an electron of spin $\sigma$ in valley $\mu$ at layer $l$ on sublattice $\alpha$ with wavevector $\bk$. The inter-valley term in the Fourier basis, describing the processes that two electrons are created in $\mu$ and $-\mu$ and get annihilated in $-\mu$ and $\mu$ valleys, reads
\begin{equation}
    \begin{split}
    \hat{V}^{\rm{inter}}&=\frac{1}{2N_s}\sum_{\alpha\alpha ', l l'}\sum_{\mu ,\sigma\sigma '}\sum_{\mathbf{k}\mathbf{k} '\mathbf{q}}V_{ll'}(\vert\mathbf{K}-\mathbf{K}'\vert) \hat{c}^{\dagger}_{\sigma \mu l \alpha}(\bk+\q)  \\
    &\times \hat{c}^{\dagger}_{\sigma' (-\mu) l' \alpha '}(\bk ' - \q) \hat{c}_{\sigma ' \mu l' \alpha '}(\bk ')\hat{c}_{\sigma (-\mu)  l \alpha}(\bk)\;,
    \end{split}
    \label{eq:V-inter}
\end{equation}
where we use $q \ll |\mathbf{K}-\mathbf{K}'|$ for Coulomb potential. Similarly, the on-site Hubbard interactions is expressed in the Fourier basis as
\begin{equation}
    \begin{split}
    \hat{V}^{\rm{Hub}}=\frac{U_0 a^2}{L_s^2 N_s}\sum_{\alpha, l,\mu}\sum_{\mathbf{k}\mathbf{k} '} & \hat{c}^{\dagger}_{\uparrow  \mu l \alpha}(\bk) \hat{c}^{\dagger}_{\downarrow \mu l \alpha}(\bk ') \\
    &\times  \hat{c}_{\downarrow \mu l \alpha}(\bk ')\hat{c}_{\uparrow \mu  l \alpha}(\bk)\;,
    \end{split}
    \label{eq:V-Hub}
\end{equation}
where $a$ is the atomic lattice constant.

In practice, the on-site interaction in moiré superlattices is typically an order of magnitude weaker than the long-range inter-site Coulomb interactions. As a result, the dominant contribution to the Coulomb energy comes from inter-site terms, and many studies therefore neglect the on-site Hubbard interaction $U_0$ in their calculations. A similar argument applies to intervalley interactions: because the Coulomb potential decays with the magnitude of momentum transfer, one generally finds $V(q) \gg V(|\mathbf{K}-\mathbf{K}'|)$ with moiré wavevector $q \ll |\mathbf{K}-\mathbf{K}'|$.

Nevertheless, both on-site and intervalley interactions have been shown to play a crucial role in resolving the competition between different isospin-polarized states in twisted $M+N$-layer graphene at integer fillings \cite{Liu-magnetic-vp-tdbg-nc22,zhang-tbmg-prl22,Li-mao-tbmg-exp-nc22,liu-1storderTDBG-prx-2023,jpliu-tbghf-prb21,zhang-liu-tbg-prl22,efetov-nature19,young-tbg-science19,sharpe-science-19,efetov-nature20}. In what follows, we explicitly develop the HF factorization for the intravalley contributions and the derivations for the remaining contributions proceed in an analogous manner.

\subsection{Projection in the low-energy band basis}
To alleviate the computational complexity, it is common practice to effectively reduce the dimensionality of the many-body Hilbert space. This reduction is justified by the fact that most non-interacting valence and conduction bands lie in energy far from the Fermi level. For low-energy physics, these valence (conduction) states can therefore be regarded as fully occupied (empty). Their occupations can thus be frozen in the many-body treatment, leaving an effective interacting problem with significantly fewer degrees of freedom. This procedure, known as projection or band truncation, requires expressing the interaction in the band basis, where single-particle states are classified according to their energy. The basis transformation between the original basis to the band basis is
\begin{equation}
\hat{c}_{\sigma\mu l \alpha}(\bk)\equiv \hat{c}_{\sigma\mu l \alpha \G}(\btk) =\sum_n C_{ \mu l \alpha \mathbf{G},n}(\btk)\,\hat{c}_{\sigma \mu,n}(\btk)\;,
\label{eq:transform}
\end{equation}
where we split the atomic wavevector $\bk$ into moir\'e wavevector $\btk$ plus moir\'e reciprocal lattice vector $\G$ and $C_{\mu l \alpha \mathbf{G},n}(\btk)$ is the expansion coefficient in the $n$-th Bloch eigenstate at $\btk$ of valley $\mu$: 
\begin{equation}
\ket{\sigma \mu, n; \btk}=\sum_{l \alpha \mathbf{G}}C_{\mu l \alpha \mathbf{G},n}(\btk)\,\ket{ \sigma \mu l \alpha \mathbf{G}; \btk }\;.
\label{eq:basis_transform}
\end{equation}

We make HF factorization such that the two-particle Hamiltonian is decomposed into a sum of the Hartree and Fock effective single-particle terms. Taking the intravalley contribution as an example, the Hartree term is expressed as
\begin{equation}
\begin{split}
\hat{V}_H^{\rm{intra}}&=\frac{1}{N_s}\sum _{\btk \btk'}\sum_{\substack{\mu\mu'\\ \sigma\sigma'\\ ll'}}\sum_{\substack{nm\\ n'm'}} \Lambda^{H}_{ \mu l nm,\mu' l' n'm'}(\btk,\btk')\\
&\times \langle \hat{c}^{\dagger}_{\sigma\mu,n}(\btk)\hat{c}_{\sigma\mu,m}(\btk)\rangle \hat{c}^{\dagger}_{\sigma'\mu',n'}(\btk')\hat{c}_{\sigma'\mu',m'}(\btk') 
\end{split}
\label{eq:intra-hartree}
\end{equation}
The Fock term is expressed as
\begin{equation}
    \begin{split}
\hat{V}_F^{\rm{intra}}&=\frac{-1}{N_s}\sum_{\btk \btk'}\sum _{\substack{\mu\mu'\\ \sigma \sigma '\\ ll'}} \sum_{\substack{nm\\ n'm'}} \Lambda^{F}_{ \mu l nm, \mu' l' n'm'}(\btk,\btk')\\
&\times \langle \hat{c}^{\dagger}_{\sigma\mu,n}(\btk')\hat{c}_{\sigma' \mu',m'}(\btk')\rangle \hat{c}^{\dagger}_{\sigma' \mu',n'}(\btk)\hat{c}_{\sigma\mu,m}(\btk)\;,   
\end{split}
\label{eq:intra-fock}
\end{equation}
where we treat spin and valley on an equal footing and the explicit form for the form factor $\Lambda^{H/F}$ can be found in literature \cite{zhang-liu-tbg-prl22,zhang-tbmg-prl22,guo-HFfci-prb-2024}. Here we assume that the translational symmetry of the moir\'e superlattice is preserved, thereby excluding the possibility of density-wave states. To study such states, one would need to explicitly enlarge the unit-cell to match the periodicity of the desired density wave and then perform HF calculations in exactly the same manner \cite{zhang-liu-tbg-prl22}. Within this framework, a projection onto a low-energy window is implemented by introducing a band-index cutoff $|n| \leq n_{\rm cut}$, where positive (negative) indices denote conduction (valence) bands relative to the charge neutrality point. Bands lying outside this cutoff are referred to as remote bands. The quadratic HF Hamiltonian to be solved self-consistently is
\begin{equation}
    \begin{split}
    \hat{H}_{\rm{HF}} = &\sum_{\substack{|n| \le n_{\rm{cut}} \\ \btk,\sigma,\mu }} \varepsilon^{0}_n (\btk) \hat{c}^{\dagger}_{\sigma\mu,n}(\btk)\hat{c}_{\sigma\mu,n}(\btk) \\
    &+ \left.  \left( \hat{V}_H^{\rm{intra}}  + \hat{V}_F^{\rm{intra}} \right) \right\vert_{|n| \le n_{\rm{cut}}}\; ,
    \end{split}
    \label{eq:HFham_general}
\end{equation}
if only intravalley contributions are considered. Here $\varepsilon^{0}_n (\btk)$ is the non-interacting band energy and the sum of Hartree and Fock part of Coulomb interactions is the HF potential.

Ideally, the band with the largest $|n|$ inside the cutoff should be separated from the remote bands by a sizable energy gap, as in TBG, where one can simply take $n_{\rm cut}=1$. In cases where such a gap is absent, it is common to choose $n_{\rm cut}=3$-$5$, providing a balance between computational efficiency and accuracy. We refer to this procedure as band-projected HF calculations.

\subsection{Workflow for self-consistent calculations}
A typical workflow for HF self-consistent calculations begins with an initial guess for the HF potential, followed by solving the HF Hamiltonian to obtain its eigenstates. From these eigenstates, one computes the mean values of the density matrix elements, which in turn define an updated HF potential. This cycle is iterated until convergence is achieved with respect to the total energy or the density matrix elements. Among all converged HF states, the one with the lowest HF total energy is identified as the HF ground state.

Although the conceptual framework of HF approximations is straightforward, their practical implementation is less trivial. This difficulty arises because the converged HF states depend sensitively on the choice of initial conditions. In principle, rigorously unconstrained HF calculations should ergodically explore the entire space of possible initial conditions, which is infinitely large. In practice, two strategies are commonly used to circumvent this challenge.

The first strategy is to generate multiple random initial HF potentials, with the hope that this sampling adequately represents the space of possible initial conditions. However, this method can easily miss the true ground state if all sampled HF potentials converge to other fixed points. While increasing the number of random initializations reduces this risk, there is no rigorous criterion to ensure that the obtained ground state is not merely metastable.

The second strategy is to construct initial HF potentials systematically by enumerating all possible order parameters associated with physical degrees of freedom, such as spin, valley, sublattice, and layer. This approach enables one to obtain all HF solutions, including both ground states and metastable states, that can be characterized by these order parameters. While it may still miss states such as density-wave states, these cases are typically treated separately. Thus, if one restricts the characterization of HF states to order parameters defined by physical degrees of freedom, this method can be regarded as an unconstrained HF calculation.

This second strategy is the one we advocate and adopt in our work. It is convenient to first specify the initial conditions in the original basis [the right-hand side of Eq.~\eqref{eq:basis_transform}], where all physical degrees of freedom are explicit, and then transform them into the band basis using Eq.~\eqref{eq:basis_transform}, with the imposed band cutoff.

\subsection{Remote band renormalization effects}
The naive projection described above completely neglects Coulomb interactions between electrons inside and outside the cutoff, i.e., the remote band effects. These interactions can significantly modify the effective low-energy non-interacting Hamiltonian, as it is well established in the case of monolayer graphene \cite{gonzalez_nuclphysb1993}. In particular, long-range exchange interactions lead to a logarithmic enhancement of the Fermi velocity of Dirac fermions. Such effects can be treated perturbatively and analytically within the framework of the renormalization group (RG). Below, we briefly outline how this procedure can be applied to twisted graphene systems and present the logarithmic corrections that arise under the RG flow. The detailed derivations are available in the literature \cite{kang-rg-prl20,guo-HFfci-prb-2024}.

Starting from
\begin{equation}
    \hat{V}_{\rm{int}} = \frac{1}{2} \int d^2 r d^2 r ' V_c(\br - \br ') \hat{\rho} (\br) \hat{\rho} (\br ') \; ,
\end{equation}
the Coulomb interactions in continuum model are defined at some high energy cut-off $E_c$. Remember that the parameters should be thought of as being fixed by a measurement at $\pm E_c$ with fully empty or occupied Hilbert space and without $e$-$e$ interactions in the framework of continuum model. Now we change $E_c$ to a slightly smaller cut-off $E_c'$ and integrate out the part outside the new cut-off, which would modify the parameters of the non-interacting Hamiltonian. This modification can be treated perturbatively when $E_c'$ is much larger than any other energy scales in the system. To do so, we split the field operator into the sum of the fast modes $\hat{\psi}^{>} (\br)$ with energy between the two energy cutoffs and the slow modes $\hat{\psi}^{<} (\br)$ with energy within $E_c'$. Then, we expand $\hat{V}_{\rm{int}}$ in terms of the two types of field operators. Heuristically, one can consider these two modes representing two different species such that $\hat{\psi}^{>} (\br)$ and $\hat{\psi}^{> \dagger} (\br)$ must appear equal times in the expansion otherwise it would vanish by taking the non-interacting mean value $\langle \dots \rangle_0$.

\begin{figure*}
    \centering
    \includegraphics[width=0.95\textwidth]{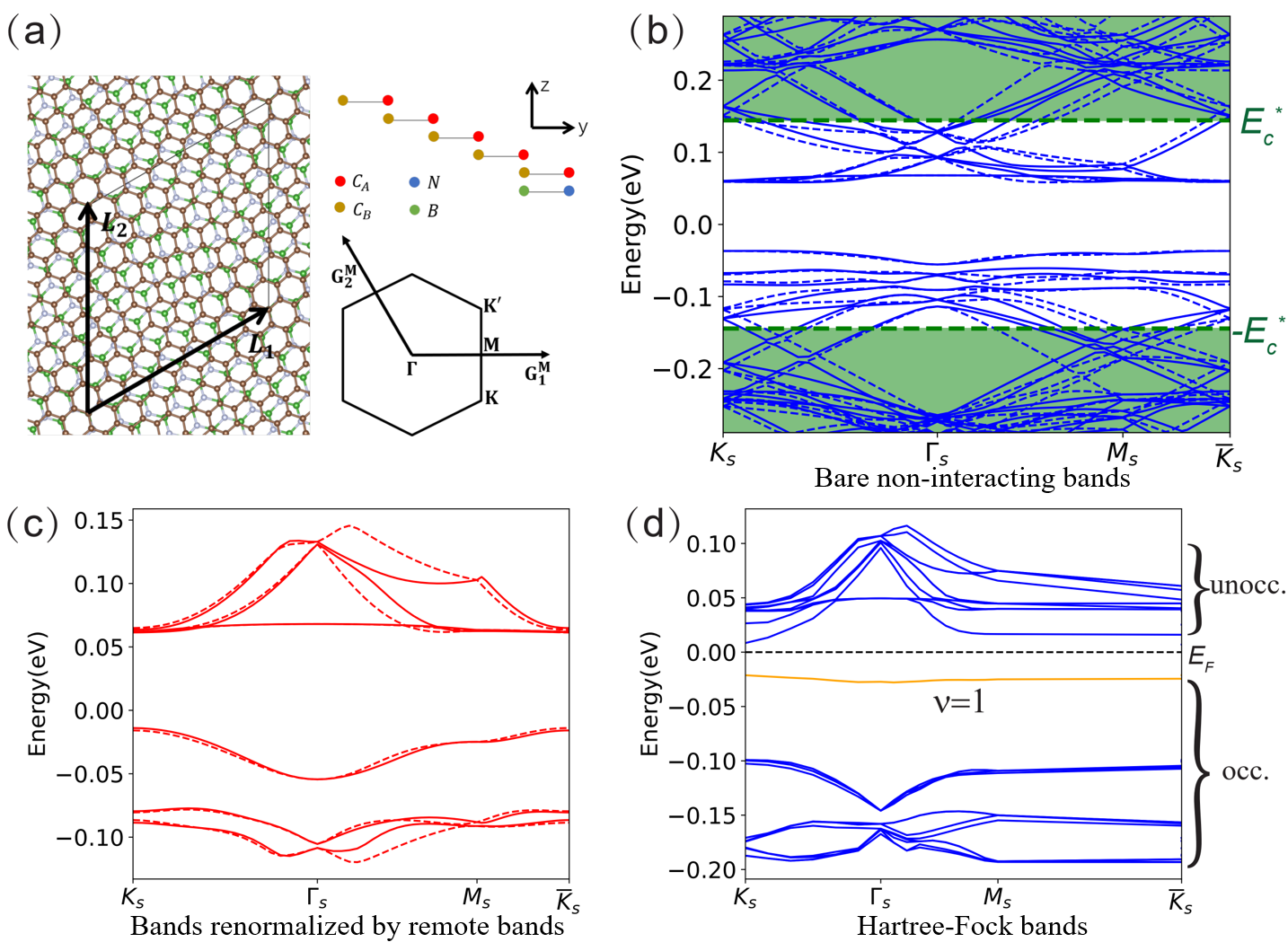}
    \caption{(a) Schematic illustration of the moir\'e superlattice consisted of nearly aligned hBN and pentalayer graphene with rhombohedral stacking. $C_A$, $C_B$ denote carbon atoms in $A$, $B$ sublattices, and $B$, $N$ denote boron and nitrogen atoms. The moir\'e Brillouin zone is plotted. (b)-(d) show continuum-model band structures of hBN-pentalayer graphene moir\'e superlattice with twist angle $0.77\,^{\circ}$ with $D=0.97\,$V/nm. (b) The bare non-interacting band structures, where the green dashed lines mark the low-energy window $E_c^*$ (see text), and the Green area denotes the remote energy bands. The solid and dashed blue lines represent the bands from $K$ and $K'$ valleys, respectively. (c) Low-energy band structures within $E_c^*$ with renormalized continuum model parameters given by Eqs.~\eqref{eq:RG_eq1}-\eqref{eq:RG_eq4}. The solid and dashed blue lines represent the bands from $K$ and $K'$ valleys, respectively. (d) Hartree-Fock band structures at filling $\nu=1$, where the gray dashed line marks the chemical potential, and the occupied and unoccupied subspaces within $E_c^*$ are marked. The orange line represents the isolated flat band right below the chemical potential of filling 1. Figure adapted from Guo et al., Phys. Rev. B, Vol.110, 075109, 2024 \cite{guo-HFfci-prb-2024}; licensed under a Creative Commons Attribution (CC BY) license.}
    \label{fig:HF_R5G_hBN}
\end{figure*}

Taking hBN-aligned rhombohedral pentalayer graphene (R5G) heterostructure as an example, whose set up is given by Fig.~\ref{fig:HF_R5G_hBN}(a), the low-energy physics of the system around $K$ or $K'$ valley can be properly described by the following non-interacting continuum model
\begin{equation}
H^{0,\mu}=H_{\textrm{R5G}}^{0,\mu}+V_{\rm{hBN}}
\end{equation}
where $H_{\textrm{R5G}}^{0,\mu}$ is the non-interacting $\k\cdot\p$ Hamiltonian of R5G of valley $\mu$ ($\mu=\pm$ standing for $K/K'$ valley), and $V_{\rm{hBN}}$ denotes the moir\'e potential exerted on the bottom graphene layer. More specifically, $H_{\textrm{R5G}}^{0,\mu}$ is comprised of the intralayer term $h_{\textrm{intra}}^{0,\mu}$ and the interlayer term $h_{\textrm{inter}}^{0,\mu}$: 
\begin{align}
&H^{0,\mu}_{\rm{R5G}}\;\nn
=&\begin{pmatrix}
h_{\rm{intra}}^{0,\mu} &(h_{\textrm{inter}}^{0,\mu})^{\dagger}  & 0 & 0 & 0 \\
h_{\textrm{inter}}^{0,\mu}& h_{\rm{intra}}^{0,\mu} & (h_{\textrm{inter}}^{0,\mu})^{\dagger} & 0 & 0\\
0 & h_{\textrm{inter}}^{0,\mu} & h_{\rm{intra}}^{0,\mu} & (h_{\textrm{inter}}^{0,\mu})^{\dagger}&0\\
 0& 0& h_{\textrm{inter}}^{0,\mu}& h_{\rm{intra}}^{0,\mu} & (h_{\textrm{inter}}^{0,\mu})^{\dagger} \\ 
 0& 0& 0 & h_{\textrm{inter}}^{0,\mu}& h_{\rm{intra}}^{0,\mu} 
\end{pmatrix}\;,
\label{eq:H0}
\end{align}
Here, the intralayer part is just the monolayer $\k\cdot\p$ model 
\begin{equation}
h_{\textrm{intra}}^{0,\mu}=-\hbar v_F^0\,\k\cdot\bm{\sigma}_{\mu}\;,
\end{equation}
where $v_F^{0}$ is the non-interacting Fermi velocity of Dirac fermions in monolayer graphene, $\bk$ is the wavevector expanded around the Dirac point from valley $\mu$, and $\bm{\sigma}_{\mu}=(\mu\sigma_x,\sigma_y)$ is the Pauli matrix defined in sublattice space. The interlayer term is expressed as
\begin{equation}
h_{\textrm{inter}}^{0,\mu}=\begin{pmatrix}
\hbar v_{\perp}(\mu k_{x}+ik_y) & t_{\perp}\;\\
\hbar v_{\perp}(\mu k_{x}-ik_y) & \hbar v_{\perp}(\mu k_{x}+ik_y) 
\end{pmatrix}
\end{equation}
where $t_{\perp}=0.34$ eV and $\hbar v_\perp=0.335$ eV$\cdot$\AA.
The moir\'e potential reads
\begin{equation}
V_{\rm{hBN}}(\vr)=V_{\rm{eff}}(\vr)+M_{\rm{eff}}(\vr)\sigma_z+ e v_F^0\,\mathbf{A}_{\rm{eff}}(\vr)\cdot\bm{\sigma}_{\mu}
\end{equation}
where $V_{\rm{eff}}$, $M_{\rm{eff}}$, and $\mathbf{A}_{\rm{eff}}$ represent the scalar moir\'e potential, Dirac mass term, and pseudo-vector-potential term, respectively.

The RG flow equations for these energy parameters in the continuum model are

    \begin{align}
        \frac{d v_F}{ d \log E_c} &= - \frac{e^2}{16 \pi \epsilon_0 \epsilon_r} \\
        \frac{d V^{\rm{eff}}}{d \log E_c} &= 0 \\
        \frac{d M^{\rm{eff}}}{d \log E_c} &= - 2 \times \frac{e^2}{16 \pi \epsilon_0 \epsilon_r v_F} \\
        \frac{d (e v_F \mathbf{A}^{\rm{eff}})}{d \log E_c} &= - \frac{e^2}{16 \pi \epsilon_0 \epsilon_r v_F} \\
        \frac{d t_\perp}{d \log E_c} &= - \frac{e^2}{16 \pi \epsilon_0 \epsilon_r v_F} \\
        \frac{d v_\perp}{d \log E_c} &= 0 \; .
        \label{eq:RG_flow}
    \end{align}

This process can be gradually done to a low-energy window $E_c^*$ until the size of Hilbert space reaches the numerical ability to do the HF calculations as long as $E_c^*$ is still the dominant energy scale in the problem, as shown in Fig.~\ref{fig:HF_R5G_hBN}(b).   

In the band-projected HF calculations, the band cut-off $n_{\text{cut}}$ serves as a natural energy cut-off for $E_c^*$. Technically, the ratio $L_s/n_{\text{cut}} a_0$ plays the role of $E_c/E_c^*$ so that

    \begin{equation}
        \frac{v_F^*}{v_F^0} =  1 + \frac{e^2}{16 \pi \epsilon_0 \epsilon_r v_F^0} \log \left( \frac{L_s}{n_{\text{cut}} a_0} \right)  \;,
        \label{eq:RG_eq1}
    \end{equation}
    \begin{equation}
        \frac{M^{\rm{eff},*}}{M^{\rm{eff}} } = \left(1 + \frac{e^2}{16 \pi \epsilon_0 \epsilon_r v_F^0} \log \left( \frac{L_s}{n_{\text{cut}} a_0} \right) \right)^2 \;,
        \label{eq:RG_eq2}
    \end{equation}
    \begin{equation}
        \frac{t_\perp^*}{t_\perp} =   + \frac{e^2}{16 \pi \epsilon_0 \epsilon_r v_F^0} \log \left( \frac{L_s}{n_{\text{cut}} a_0} \right)   \;,
        \label{eq:RG_eq3}
    \end{equation}
    \begin{equation}
        \frac{e v_F^* \mathbf{A}^{\rm{eff},*}}{e v_F^0 \mathbf{A}^{\rm{eff}}} =  1 + \frac{e^2}{16 \pi \epsilon_0 \epsilon_r v_F^0} \log \left( \frac{L_s}{n_{\text{cut}} a_0} \right) \; .
        \label{eq:RG_eq4}
    \end{equation}

A typical RG non-interacting band structure is shown in Fig.~\ref{fig:HF_R5G_hBN}(c). Note that most of these parameters are enhanced logarithmically, in particular the Dirac velocity $v_F$, while the energy scale of Coulomb interactions is fixed under the RG process, which is given by the average distance between electrons. Therefore, the effective strength of Coulomb interactions is weakened under RG process.

It is important to note that this renormalization effect arises only for Dirac Hamiltonians with terms linear in $\bk$. Consequently, tTMDs do not exhibit such an effect, since their long-wavelength expansion is parabolic in $\bk$. This observation further justifies neglecting electrons excluded from the continuum model, as their non-interacting dynamics are not governed by a Dirac Hamiltonian. Nevertheless, one should in principle still account for remote band screening, as we discuss next.

\subsection{Remote band screening: constrained random phase approxiamtion}
The Coulomb interaction between two electrons occupying the low-energy bands can also be screened due to the virtual excitations of particle-hole pairs from the remote bands. We consider the screening effects to the low-energy window electrons from the virtual excitations of particle-hole pairs through the following three processes: the particle-hole pairs are excited (\i) from the remote bands below charge neutrality point (CNP) to those above CNP, (\ii) from remote bands below CNP to the low-energy bands, and (\iii) from the low-energy bands to remote bands above CNP. However, we exclude the particle-hole excitations within the band-cutoff window. This is called the constrained random phase approximation (cRPA) \cite{pollet-tbg-crpa-prb20,wehling-tbg-prb19,zhang-liu-tbg-prl22,fqah-luxb-natmat25}.

In the band basis, this amounts to replace the diagonal bare interaction $V$ by the cRPA screened Coulomb interaction matrix $V^{\textrm{cRPA}}(\btq)_{\mathbf{Q},\mathbf{Q}^{\prime}}$ given by
\begin{align}
    \left[V^{\rm{cRPA}}(\btq)\right] = \left[V(\btq+\Q)\delta_{\mathbf{Q},\mathbf{Q}'}\right] \left[\epsilon^{\rm cRPA}(\btq)\right]^{-1}
\end{align}
where the bare interaction is only screened by the environment dielectric constant $\epsilon_{\rm env}$ and the cRPA dielectric matrix
\begin{equation}
\epsilon^{\rm cRPA}_{\mathbf{Q},\mathbf{Q}^{\prime}}(\btq)=\left(1+ \left[\chi_c^{0}(\btq) \right] \left[V(\btq+\Q)\delta_{\mathbf{Q},\mathbf{Q}'} \right]  \right)_{\mathbf{Q},\mathbf{Q}^{\prime}}
\label{eq:epsilon}
\end{equation}
where the brackets are to emphasize the matrix form of the quantities. Here, the constrained susceptibility $\chi_c^{0}(\btq) $ is only to sum up the three processes of the particle-hole excitations that are described above. We further project the cRPA screened Coulomb interactions onto the low-energy bands and then carry out the HF calculations as described before. 

\begin{figure}
    \centering
    \includegraphics[width=0.45\textwidth]{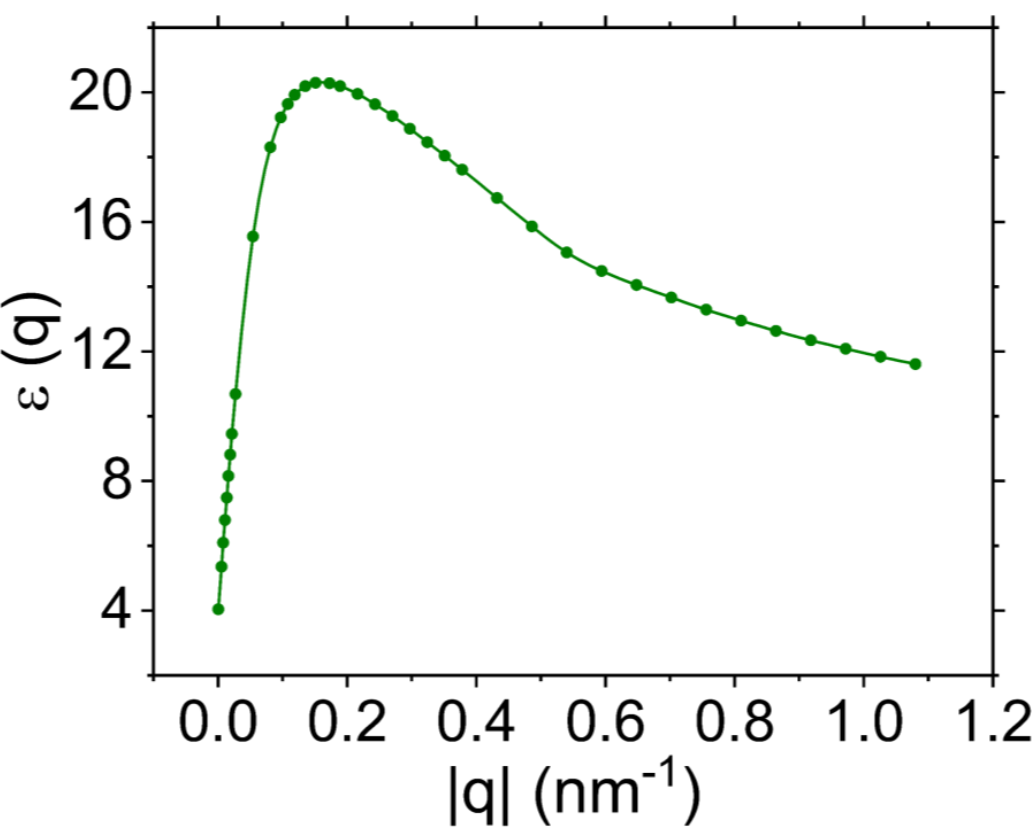}
    \caption{The wavevector dependence of the cRPA dielectric constant calculated for TBG. Figure extracted from Zhang et al., Phys. Rev. Lett. Vol.128, 247402, 2022 \cite{zhang-liu-tbg-prl22}; licensed under a Creative Commons Attribution (CC BY) license.}
    \label{fig:epsiloncRPA}
\end{figure}

In practice, the dielectric matrix can be simplified to its dominant diagonal elements, whose tyical dependence on $\btq$ is shown in Fig.~\ref{fig:epsiloncRPA}. A pronounced peak structure at specific wavevectors indicates that the interaction channel at those $\btq$ values is strongly suppressed compared to other momentum transfers. This leads to inhomogeneous screening within the moir\'e Brillouin zone, which cannot be captured by a single phenomenological dielectric constant $\epsilon_r$. Such effects may become particularly important when the HF band dispersion serves as input for subsequent calculations.

Unlike the remote-band renormalization effect, however, cRPA screening is relevant for both twisted graphene systems and tTMDs, representing a physically different mechanism. In practice, both effects can be incorporated simultaneously: one first applies the RG procedure using the environmental dielectric constant in Eqs.~\eqref{eq:RG_eq1}-\eqref{eq:RG_eq4}, and then computes the cRPA dielectric matrix to obtain the screened Coulomb interactions used in the HF self-consistent calculations.

\subsection{Normal ordering}
There is no unique prescription for normal-ordering the Coulomb interactions in band-projected HF calculations, and different choices can lead to qualitatively distinct results \cite{kwan-RnGHF-prb-2025,guo-HFfci-prb-2024,dong-R5GAHC-prl-2024,dong-R5Gfci-prl-2024,zhou-R5Gfci-prl-2024}. Therefore, it is important to discuss the various normal-ordering schemes employed in the literature. Here, we take the hBN-R5G heterostructure as an example, since different groups have applied different normal-ordering conventions to the same system. While we cannot claim that any single normal-ordering scheme is universally superior, it is clear that one choice is unlikely to be accurate to describe hBN-R5G heterostructures.

\subsubsection{Normal ordering with respect to low-energy Hilbert-space vacuum}
In our framework \cite{guo-HFfci-prb-2024}, the $e$-$e$ interactions is normal-ordered with respect to the vacuum of the truncated low-energy Hilbert space. To account for remote band effects, we first integrate out the fast modes associated with remote bands, yielding a low-energy effective model with cutoff $E_c^*$. The $e$-$e$ interactions are then projected onto the ``non-interacting'' wavefunctions of this renormalized low-energy model, which captures the effective single-particle behavior within the low-energy window. Although the $e$-$e$ interactions within this window have not yet been included, the influence of the remote band electrons has already been incorporated via the RG procedure. Consequently, the $e$-$e$ interactions should be normal-ordered with respect to the vacuum of the truncated renormalized low-energy Hilbert space.

\subsubsection{Normal ordering with respect to filled bands at charge neutrality}

One can also consider another type of normal ordering with respect to CNP and consider the valence band electrons are also spectator. This amounts to subtract the HF density matrix $\langle \hat{c}^{\dagger}_{\sigma\mu,n}(\kt)\hat{c}_{\sigma\mu',n'}(\kt)\rangle$ by the non-interacting density matrix $\rho^0_{\sigma\mu n, \sigma\mu' n'}(\kt)$:
\begin{equation}
\langle \hat{c}^{\dagger}_{\sigma\mu,n}(\kt)\hat{c}_{\sigma\mu',n'}\rangle\to
\langle \hat{c}^{\dagger}_{\sigma\mu,n}(\kt)\hat{c}_{\sigma\mu',n'}\rangle-\rho^0_{\sigma\mu n, \sigma\mu' n'}(\kt)\;,
\label{eq:no-ii}
\end{equation} 
with 
\begin{equation}
\rho^0_{\sigma\mu n, \sigma\mu' n'}(\kt)=
\begin{cases}\delta_{\mu\mu'}\delta_{nn'}\;, \hspace{12pt}  \textrm{ if $\epsilon_{\mu n}(\kt)\in$ occupied}\\
0\;, \hspace{12pt} \textrm{ if $\epsilon_{\mu n}(\kt)\in$ unoccupied}
\end{cases}
\end{equation}
where $\epsilon_{\mu n}(\kt)$ is the non-interacting band energy of the renormalized low-energy continuum model at moir\'e wavevector $\kt$, $\sigma$ and $\mu$ are the spin and valley indices. 

It should be noted that adopting such a normal ordering (Eq.~\eqref{eq:no-ii}) implicitly assumes that the valence and remote bands are completely omitted, which is not physically realistic. As we have already discussed, remote bands play an important role in both renormalization effects and cRPA screening. Similarly, the valence bands can significantly influence the effective low-energy Hamiltonian. Occupying the valence bands generates a background charge density arising solely from these states \cite{kwan-RnGHF-prb-2025}. This charge density generates a background moir\'e potential and modifies the HF single-particle spectrum.

\subsubsection{``Infinite-temperature'' normal ordering within low-energy Hilbert space}
Alternatively, one can also consider normal ordering $e$-$e$ Coulomb interactions with consistent convention for all energy scales. In reciprocal space, such an interaction is expressed as 
\begin{equation}
\hat{V}_{\rm{int}}=\frac{N_s}{2}\sum_{\mathbf{q}}V(\mathbf{q})\delta \hat{\rho}(\mathbf{q})\delta \hat{\rho}(-\mathbf{q}),
\label{eq:Vq-drho-drho}
\end{equation}
where $\delta \hat{\rho}(\q)$ is defined as
\begin{equation}
   \delta\hat{\rho}(\mathbf{q})=\frac{1}{N_s}\sum_{\bk}\sum_{\mu\sigma\alpha}\left(\,\hcd_{\sigma\mu l \alpha}(\bk+\q)\hc_{\sigma\mu l \alpha}(\bk)-\frac{1}{2} \delta_{\q,\mathbf{0}}\right).  
\end{equation}
Compared with the $e$-$e$ interaction term that is normal ordered with respect to the vacuum of the low-energy Hilbert space, Eq.~\eqref{eq:Vq-drho-drho} would introduce additional $\bk$ dependent single-particle terms, which can be interpreted as the remote band potential. This remote band potential is even exact if the particle-hole symmetry is present \cite{Bernevig-tbg3-arxiv20}. Therefore, such kind of normal ordering is also widely adopted in studying correlated states in magic-angle TBG. We can also use this normal ordering when particle-hole symmetry breaking terms is small compared to the energy cutoff $E_c^*$ \cite{kang-rg-prl20,Daniel-Ashvin-fci-tbg-arxiv21,zhang-liu-tbg-prl22,kwan-RnGHF-prb-2025}.

\subsection{All-band HF calculations}
Calculations show that our normal ordering (the first scheme) produces a HF phase diagram qualitatively similar to that obtained using the ``infinite-temperature'' normal ordering for the hBN-R5G heterostructure at $\theta = 0.77^\circ$ and moir\'e filling $\nu = 1$ \cite{fqah-julong-nature24,guo-HFfci-prb-2024,kwan-RnGHF-prb-2025}. Most saliently, both schemes reproduce qualitatively the measured phase diagram. In contrast, the phase diagram obtained using the second normal-ordering scheme differs significantly from those of the first and third schemes, highlighting the important roles of remote-band and valence-band electrons. 

The arbitrariness in normal ordering stem from the fact that the band truncation used in the band-projected HF calculations introduces uncontrolled errors, as the exchange interaction with high-energy bands remains significant, especially when the band gaps are smaller than the typical Coulomb interaction strength. To address this issue, one can perform all-band HF calculations, in which all bands within the plane-wave cutoff of the continuum model are included in the self-consistent cycles. This approach eliminates any arbitrariness in normal ordering since no projection is involved and the problem is automatically solved. Core electrons and valence electrons beyond the plane-wave cutoff (typically on the order of  $\sim$eV) are no longer described by the Dirac equation and act merely as spectators within the continuum model framework. Consequently, remote-band effects are automatically incorporated, and the vacuum is defined as the set of empty bands of the non-interacting continuum model.

Due to their high computational cost, all-band HF calculations have only recently been applied to the hBN-R5G heterostructure \cite{lu-MBPTmoire-arxiv-2025}, where improved agreement with the experimental results \cite{fqah-julong-nature24} has been achieved. All-band HF calculations are particularly indispensable for many-body perturbation theory, where convergence with respect to the band cutoff is slow, making band-projected HF irrelevant for the subsequent perturbation theory. Further details will be provided in the following section.

\subsection{Limitations in HF approximations}
Although HF approximations have been successful in explaining experimental observations and predicting correlated phases in twisted systems, they have inherent limitations. First, HF accounts only for exchange effects between electrons, neglecting all correlation effects. Correlations are the origin of screening that reduces the bare Coulomb potential, which means HF tends to favor symmetry-breaking states, such as correlated insulators. As a result, some states predicted by HF may not be realized experimentally. Conversely, it is often easier to use HF to reproduce phases that are already observed in experiments. While several methods exist to partially account for screening, they are generally static and homogeneous across the Brillouin zones. Inhomogeneous and dynamic screening effects remain unaddressed. As a result, HF approximations miss both the energy-dependent softening and the non-local nature of screening, leading to inaccurate quasiparticle spectrum.

Another limitation is that HF is a mean-field approach, which restricts all resulting correlated states to Slater determinants. This is insufficient to capture strongly correlated phases, such as fractional Chern insulators. These two challenges for HF approximations will be addressed by many-body perturbation theory and exact diagonalization technique, respectively, in the following sections.

\section{Many-body perturbation theory}
Many-body perturbation theory (MBPT) takes a step beyond HF calculations by incorporating correlation effects perturbatively. While most textbooks introduce perturbation theory starting from a non-interacting state and treating interactions diagrammatically \cite{fetter-book}, it is important to emphasize that the concept of perturbation is more general. The only requirement for perturbation theory to be valid is that no phase transition occurs upon inclusion of the perturbation. In other words, the perturbed state must be adiabatically connected to the unperturbed state, which typically means no change in symmetry or topological properties. Therefore, as long as HF states capture the qualitative properties observed experimentally—which is often the case—MBPT can improve the accuracy of theoretical predictions. In practice, perturbation theory can also be used a posteriori to assess the validity of HF states: if perturbative corrections only quantitatively modify the starting states without inducing qualitative changes (e.g., gap closure), the HF states are a good description of reality; otherwise, if a phase transition occurs, they are not.

This section focuses on two key applications of MBPT in many-electron systems. First, we show how correlation effects—completely neglected in HF—modify the total energy of a given HF state using the RPA. This helps identify the true many-body ground state among different converged HF states, yielding a phase diagram that better agrees with experiments. Second, we apply MBPT to the self-energy using the $GW$ approximation, which accounts for inhomogeneous screening and dynamical effects in the dielectric matrix. This procedure generally corrects HF band structures to the $GW$ quasiparticle band structures. The differences arise from the real part of the self-energy, while the imaginary part, representing quasiparticle lifetimes, provides an estimate of the strength of correlations in the system.

\subsection{Random phase approximation for correlation energy}
The total energy of the system within RPA framework is given by:
\begin{equation}
	E_{\text{tot}}=E_{\text{kin}}+E_{\text{HF}}+E_{c}^{\text{RPA}},
\end{equation}
where $E_{\text{kin}}$ is the kinetic energy, $E_{\text{HF}}$ is the HF energy, and $E_{c}^{\text{RPA}}$ represents the correlation energy obtained through the RPA. The inclusion of (negative) $E_{c}^{\text{RPA}}$ is crucial, as the HF approximation alone neglects correlation effects, thus biased into symmetry-breaking states. The RPA provides a more accurate description by incorporating the effects of $e$-$e$ interactions beyond the mean-field level.

\begin{figure}
    \centering
    \includegraphics[width=0.45\textwidth]{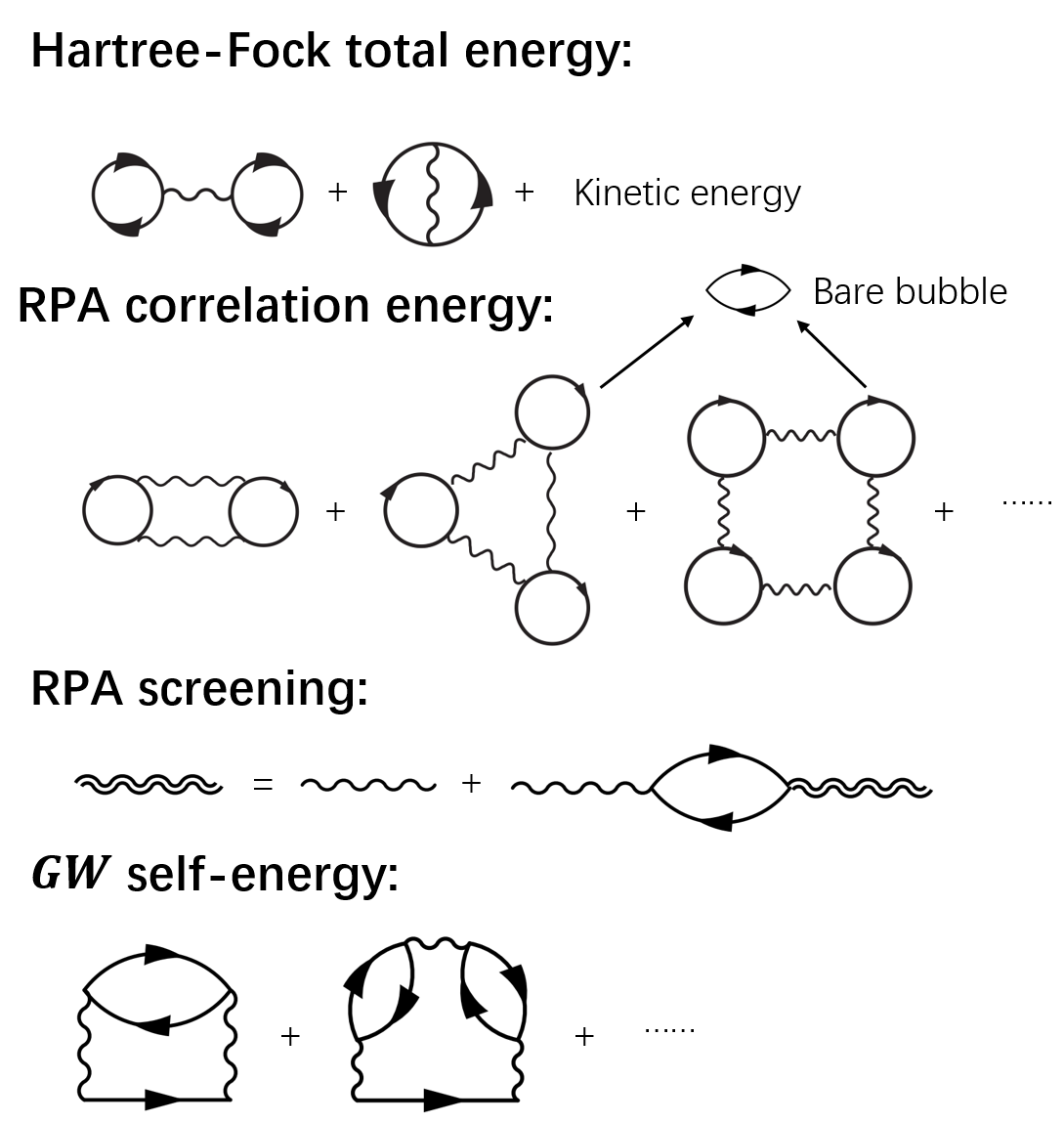}
    \caption{Feynman digrams that are considered in HF total energy, RPA correlation energy $E_{c}^{\text{RPA}}$, RPA screened Coulomb potential $W$ and $GW$ self-energy $\Sigma$. Single wiggling lines represents bare Coulomb interaction, double wiggling line is the RPA screened Coulomb interactions and solid lines with an arrow are the bare Green's function.}
    \label{fig:GWRPAdiagram}
\end{figure}

Diagrammatically, $E_{c}^{\text{RPA}}$ includes the sum of all the ring diagrams that can be written in the form of two rings only connected by a bare interaction line and an RPA-screened interaction line that considers only bubble diagrams for electron-hole fluctuations, as shown in Fig.~\ref{fig:GWRPAdiagram}. This would include the contribution from plasmonic collective excitations to the total energy. It also considers the frequency dependent screening that cannot be taken into account by, for example, the use of phenomenological dielectric constant or the cRPA method. Mathematically, the correlation energy in the RPA is given by \cite{fetter-book,bohm-rpa-pr53,gellmann-rpa-pr57,ren-rpa-2012}:
\begin{equation}
	\begin{aligned}
		E_c^{\text{RPA}}&=\int_{-\infty}^{\infty} \frac{\mathrm{d} \omega}{4 \pi} \, \mathrm{Tr} \left\{ \ln \left[1-V\chi^0(i\omega)\right] + V\chi^0(i\omega) \right\} \\
		&=\int_{-\infty}^{\infty} \frac{\mathrm{d} \omega}{4 \pi}  \sum_{\btq,\Q,\Q'} \left\{ \ln \left[\delta_{\Q\Q'} - V_{\Q }(\btq)\delta_{\Q\Q'}\chi^0_{\Q'\Q }(\btq,i\omega)\right] \right. \\
        &\hspace{2cm} \left. + V_{\Q }(\btq)\delta_{\Q\Q'}\chi^0_{\Q'\Q }(\btq,i\omega) \right\},
	\end{aligned}
    \label{eq:EcRPA}
\end{equation}
where $V_{\Q }(\btq)\equiv V(\Q + \btq)$ is the bare Coulomb interaction, and $\chi^0(i\omega)$ is the bare charge polarizability in the imaginary frequency domain. It should be emphasized that we call $\chi^0$ the bare charge polarizability instead of the non-interacting one since it is still well-defined if the state is still a Slater determinant such for HF states. 

The first line of the equation presents a general expression for the RPA correlation energy in matrix form, involving a trace over all possible interaction channels. The second line expands this expression into the momentum space for general moir\'e systems, where $\btq$ and $\Q$ denote momentum vectors in the moir\'e Brillouin zone and moir\'e reciprocal lattice vectors, respectively. The matrix elements of the zero-temperature bare charge susceptibility $\chi^0$ in moir\'e reciprocal space can be explicitly written as
\begin{equation}
	\begin{aligned}
		\chi^{0}_{\Q\Q'}(\btq,\nu)&=\frac{1}{N_s \Omega_0}\sum_{n',n,\btk}
		\Lambda^*_{n'n}(\btq,\Q) \Lambda_{n'n}(\btq,\Q')\\
		&\times\left[\frac{\theta(\varepsilon_{n'\btk+\btq}-\varepsilon_F)\theta(\varepsilon_F-\varepsilon_{n\btk})}{\nu-\varepsilon_{n'\btk+\btq}+\varepsilon_{n\btk}+i\delta} \right.\\
		&\left. -\frac{\theta(\varepsilon_F-\varepsilon_{n'\btk+\btq})\theta(\varepsilon_{n\btk}-\varepsilon_F)}{\nu-\varepsilon_{n'\btk+\btq}+\varepsilon_{n\btk}-i\delta}\right],
        \label{eq:chi0}
	\end{aligned}
\end{equation}
where $N_s$ is the total number of moir\'e supercells, $\Omega_0$ represents the area of the moir\'e unit-cell. Here, the form factor $\Lambda_{nn'}$ between two moir\'e bands $n$ and $n'$ can be expressed using the expansion coefficients $C_{\lambda\G, n\btk}$ of the single-particle states in the plane-wave basis as 
\begin{equation}
    \Lambda_{n'n}(\btq,\Q) = \sum_{\lambda,\G} C^*_{\lambda \G+\Q, n'\btk+\btq} C_{\lambda \G, n\btk}
\end{equation}
where $\lambda$ encodes all the other degrees of freedom such as spin, valley, sublattice and layer, and $\btk$ and $\G$ denote wave vectors in the moir\'e Brillouin zone and moir\'e reciprocal lattice vectors, respectively.

The numerical computation of $E_{c}^{\text{RPA}}$ is preferably performed on the imaginary axis, as shown explicitly in Eq.~\eqref{eq:EcRPA}, which facilitates numerical convergence because the poles of the integrand, determined by $\chi^0$, lie close to the real axis. However, $\chi^0$ usually exhibits a kink at $\omega=0$, where the imaginary axis intersects the real axis. It is therefore advantageous to split the integral at $\omega=0$ and compute the two parts separately, which significantly accelerates convergence. Calculations also show that $E_{c}^{\text{RPA}}$ converges slowly with respect to the number of bands included in $\chi^0$. Consequently, imposing a band cutoff is not acceptable. The most reliable approach is to include all bands within the plane-wave cutoff of the continuum model, making all-band HF calculations essential. This is physically reasonable, provided the assumption holds that electrons beyond the plane-wave cutoff act as spectators.

Although $E_{c}^{\text{RPA}}$ does not account for ladder diagrams representing higher-order exchange effects such as excitonic excitations, it already provides a substantial improvement over HF results. For example, within the HF approximation, the Wigner crystal transition occurs at a critical $r_s \sim 2.2$, far below the experimental value of $35.1(9)$ \cite{guo-AHC-arxiv-2025}. Including the RPA correlation energy shifts the Wigner crystal transition to $r_s \sim 3.9$, roughly twice the HF value \cite{guo-AHC-arxiv-2025}. While a considerable gap remains due to the strongly correlated nature of the Wigner crystal, this improvement highlights the stabilizing role of correlations in the Fermi liquid state. This can be understood from the denominator in Eq.~\eqref{eq:chi0}, where intraband transitions contribute more significantly than interband transitions. A practical rule of thumb is that RPA correlation energy tends to favor metallic states over insulating states, and small-gap insulators over large-gap insulators.

A recent successful application of RPA correlation energy in moir\'e systems is the interpretation of transport measurements in hBN-R5G across a wide range of vertical electric fields \cite{lu-MBPTmoire-arxiv-2025}. Previous HF calculations tended to overestimate the stability of gapped states, whereas including RPA corrections yields quantitative agreement with experiments.

\subsection{$GW$ approximation}
The $GW$ approximation represents a series type of approximation for incorporating screening effects into self-energy beyond the mean-field HF theory \cite{hedin-gw-pr65,louie-ppa-prb86,aryasetiawan-gw-rpp98,rubio-rmp02,reining-gw-wires18,golze-gw-fc19}. It improves the description of quasiparticle energies and the single-particle energy spectrum by including the effects of dynamical screening. Historically, $GW$ approximation is the first attempt to simplify a set of mathematically exact equations, known as Hedin's equations \cite{hedin-gw-pr65}, which describe a set of self-consistent equations of self-energy $\Sigma$, single-particle Green's function $G$, vertex function $\Gamma$, polarization propagator $\chi$, and screened Coulomb interaction $W$. For notational simplicity, here we use Arabic number as a shorthand notation for spatial ($\br$) and temporal ($t$) coordinate, e.g., $1\equiv(\br_1, t_1)$. Then, with such shorthand notations, Hedin's equations are given by

\begin{equation}
	\Sigma(12) = i \int \text{d}3 \, G(13) W(14) \Gamma(342),
\end{equation}
\begin{equation}
	G(12) = G_0(12) + \int \text{d}3 \, G_0(13) \Sigma(34) G(42),
\end{equation}
\begin{equation}
	\begin{split}
    \Gamma(123) =\delta(12) \delta(13) + &\int \text{d}4 \, \text{d}5 \, \frac{\delta \Sigma(12)}{\delta G(45)} \\
    & \times G(46) G(75) \Gamma(673),
    \end{split}
\end{equation}
\begin{equation}
	\chi(12) = -i \int \text{d}3 \, \text{d}4 \, G(13) G(42) \Gamma(342),
\end{equation}
\begin{equation}
	W(12) = V(12) + \int \text{d}3 \, V(13) \chi(34) W(42).
\end{equation}

Hedin's equations link the Green's function $G$, self-energy $\Sigma$, and screened interaction $W$ in a self-consistent way. Their derivations can be found in several publications \cite{hedin-gw-pr65,rubio-rmp02,giustino-rmp-2017}.

\subsubsection{General formalism with application to moir\'e systems}
We will apply the $GW$ approximation upon the HF calculations. Let us first assume a simple form of vertex function, $\Gamma(123) = \delta(12) \delta(13)$, which leads to the ``bare vertex'' approximation (as first did by Hedin \cite{hedin-gw-pr65})
\begin{equation}
	\Gamma(123) = \delta(12) \delta(13).
\end{equation}
Under this approximation, the self-energy simplifies to
\begin{equation}
	\Sigma(12) = i G(12) W(12) \; ,
\end{equation}
which gives the name of $GW$ approximations to self-energy. Similarly, the polarization reduces to a bubble diagram
\begin{equation}
	\chi(12) = -i G(12) G(21).
    \label{eq:polarization_bubble}
\end{equation}
To describe the screened Coulomb interaction $W$ in terms of the dielectric function $\epsilon$, we express $V$ as:
\begin{equation}
	V(12)=\int\text{d}{3}\epsilon(13)W(32),
    \label{eq:screened_W}
\end{equation}
where the dielectric function $\epsilon$ can be formulated as
\begin{equation}
	\epsilon(12)=\delta(12)-\int\text{d}{3}V(13)\chi(32).
    \label{eq:dielectric_rpa}
\end{equation}
Note that the Green's function $G$ appearing in these equations is the interacting one which is a priori unknown. Nevertheless, it can be self-consistently determined by starting from the bare Green's function $G_0$ given by
\begin{align}
	G_0(\br,\br',\omega)=\sum_{n\btk}\frac{\psi_{n\btk}(\br)\psi^*_{n\btk}(\br')}{\omega-\varepsilon_{n\btk}+i\delta_{n\btk}},
\end{align}
where $\psi_{n\btk}$ would be the HF single-particle wave functions for moir\'e miniband $n$, $\varepsilon_{n\btk}$ are the HF eigenvalues, and $\delta_{n\btk}\equiv\delta \times \mathrm{sgn}(\varepsilon_{n\btk}-\varepsilon_F)$ with the Fermi energy $\varepsilon_F$ and $\delta>0$. Here we have performed a Fourier transform from time domain to frequency domain to handle the time-dependent components and restored the usual notations where $\br$ ($\br'$) denote real-space coordinate, and $\omega$ denote frequency. The bare charge polarizability $\chi_0$ by Eq.~\eqref{eq:polarization_bubble} is given by:
\begin{equation}
	\chi_{0}(\br,\br',\nu)=-i\int\frac{\text{d}{\omega}}{2\pi}e^{i\omega 0^+}G_0(\br,\br',\omega+\nu)G_0(\br',\br,\omega).
\end{equation}
By Eq.~\eqref{eq:dielectric_rpa}, the dielectric function is expressed as
\begin{equation}
	\epsilon_{\text{RPA}}(\br, \br', \omega)=\delta(\br, \br')-\int\text{d}\br''V(\br, \br'')\chi_{0}(\br'', \br',\omega),
\end{equation}
which gives rise to the usual RPA dielectric function. Then, by Eq.~\eqref{eq:screened_W}, the RPA-screened Coulomb interaction $W_0$ reads 
\begin{equation}
	W_{0}(\br, \br', \omega) = \int \text{d}\br'' \epsilon^{-1}_{\text{RPA}}(\br, \br'', \omega) V(\br'', \br'),
\end{equation}
leading to the so called $G_0 W_0$ self-energy in literature as self-explained by 
\begin{equation}
	\Sigma(\br, \br', \omega) = \frac{i}{2\pi} \int d\nu \, e^{i\nu 0^+} G_0(\br, \br', \omega + \nu) W_{0}(\br', \br, \nu).
\end{equation}

Since a part of $GW$ self-energy, namely the Fock self-energy $\Sigma_{\rm{Fock}} = i G_0 V$, has already been included while solving the HF self-consistent equations, we are left with the correlation self-energy $\Sigma_c$ that arises from the screening potential (not ``screened''), defined as the difference between the bare Coulomb potential $V$ and the RPA-screened $W_0$
\begin{equation}
	W_{c}(\br, \br', \omega) = \int \text{d}\br'' \left[ \epsilon^{-1}_{\text{RPA}}(\br, \br'', \omega) - \delta(\br, \br'') \right] V(\br'', \br').
\end{equation}
Note that the static and homogeneous part has been already taken into account in the HF calculations via phenomenological dielectric constant, the only fitting parameter in our theory. Then, $\Sigma_c$, which accounts for electron correlation effects beyond HF, is then computed using
\begin{equation}
	\Sigma_c(\br, \br', \omega) = \frac{i}{2\pi} \int d\nu \, e^{i\nu 0^+} G_0(\br, \br', \omega + \nu) W_{c}(\br', \br, \nu),
\end{equation}

Let us give the explicit form of these quantities in the context of moir\'e systems. The matrix form of the RPA dielectric function in reciprocal space is expressed as
\begin{equation}
	\epsilon^{\text{RPA}}_{\Q\Q'}(\btq,\omega)	=\delta_{\Q\Q'}-V(\btq+\Q)\chi^{0}_{\Q\Q'}(\btq,\omega),
\end{equation}
and the screening potential in reciprocal space is given by
\begin{equation}
	W^{c}_{\Q\Q'}(\btq,\omega)=\left[ \epsilon^{-1,\text{RPA}}_{\Q\Q'} (\btq,\omega)-\delta_{\Q\Q'}\right]V(\btq+\Q').
\end{equation}
The time-ordered correlation self-energy in reciprocal space can then be expressed in the band basis
\begin{equation}
	\begin{aligned}
		&\Sigma_{c}(\btk,\omega)_{n'n}
		=\frac{i}{N\Omega_0}\sum_{m,\btq}\sum_{\Q,\Q'} \Lambda^*_{mn'}(\btq,\Q) \Lambda_{mn}(\btq,\Q')   \\
		&\times V(\btq+\Q)
		\int\frac{\text{d}{\nu}}{2\pi}
		\frac{e^{i\nu 0^+} \left[ \epsilon^{-1,\text{RPA}}_{\Q \Q'}(\btq,\nu)-\delta_{\Q\Q'}\right] }{\omega+\nu-\varepsilon_{m\btk+\btq}+i\delta_{m\btk+\btq}} \; .
	\end{aligned}
    \label{eq:selfenergy_corr}
\end{equation}
The inverse dielectric function $\epsilon^{-1}_{\text{RPA}}$ appearing in the above equation captures the frequency-dependent dynamical screening of the Coulomb interaction. Its off-diagonal components account for inhomogeneous screening across different Brillouin zones. These effects are not included in the cRPA treatment used in the HF calculations discussed in the previous section.

\subsubsection{Quasiparticle energy}
The quasiparticle energy within the $GW$ approximation can be obtained by self-consistently solving the sum of the HF Hamiltonian and the correlation self-energy $\Sigma_c$. However, as shown in Eq.~\eqref{eq:selfenergy_corr}, the self-energy operator is generally nonlocal, energy-dependent, and non-Hermitian, so the quasiparticle energy is, in general, complex. The imaginary part of the energy determines the quasiparticle lifetime, which is why we refer to it as a quasiparticle rather than a well-defined real particle. When the imaginary part is negligible, the quasiparticle closely resembles a real particle with the same dynamics. This physical picture, first introduced by Landau, can be understood as an electron dressed with a cloud of excitations arising from screening in the solid, including electron-hole pairs and plasmons.

There are different schemes to arrive at the quasiparticle energy. The simplest one is to consider only the diagonal part of $\Sigma_c$. The quasiparticle energies can be then obtained using linear expansion in $G_0W_0$ self-energy
\begin{equation}
	\varepsilon_{n\btk}^{\text{QP}} = \varepsilon_{n\btk}^{\text{HF}} + Z_{n\btk} \, \text{Re} \, \Sigma_c(\btk, \varepsilon_{n\btk}^{\text{HF}})_{nn},
    \label{eq:qp_diagonal}
\end{equation}
where $Z_{n\btk}$ is the quasiparticle weight, accounting for interaction renormalization effects of quasiparticles
\begin{equation}
	Z_{n\btk} = \left[ 1 - \text{Re} \left( \frac{\partial \Sigma_c(\btk, \omega)_{nn}}{\partial \omega} \right)_{\omega = \varepsilon_{n\btk}^{\text{HF}}} \right]^{-1}.
\end{equation}
The fact that $Z$ is close to unity suggests that using the lowest-order term in a perturbation series, i.e., the $GW$ approximation, is reasonable. This also indicates that HF approximations are not too far from reality. A one-shot $GW$ calculation, commonly referred to as the $G_0W_0$ scheme \cite{aryasetiawan-gw-rpp98,hybertsen-GW0-prb-1986}, often provides significant corrections to quasiparticle energies. A relatively inexpensive improvement is to update the HF energies in Eq.~\eqref{eq:qp_diagonal} with the quasiparticle energies and iterate until convergence. This scheme, known as EV-$GW$ (eigenvalue-only $GW$) \cite{surh-EVGW-prb-1991}, systematically improves quasiparticle energies compared to $G_0W_0$, especially when the initial HF energies are far from the final quasiparticle energies. This partly weaken the starting-point dependence of $G_0W_0$.

More sophisticated schemes exist. For example, the off-diagonal part of the self-energy can be partially included by taking its Hermitian component, as done in the $QPGW$ \cite{schilfgaarde-QPGW-prl-2006,kotani-QPGW-prb-2007} and $scGW$ \cite{shishikin-scGW-prl-2007} schemes. These approaches also address the starting-point dependence inherent in $G_0W_0$. Full self-consistent $GW$ calculations are also possible by updating both the eigenvalues and wavefunctions of the Green's function in Hedin's equations. Further refinements, such as the inclusion of vertex corrections ($GW\Gamma$ \cite{delsole-GWGamma-prb-1994,bruneval-GWfxc-prl-2005}), require a suitable exchange-correlation functional to make computations tractable.

However, these advanced schemes are not essential at this stage for several reasons. First, more sophisticated methods do not necessarily yield better results, due to cancellation effects between explicit vertex corrections in $\Gamma$ and implicit ones in $W$ \cite{kotani-QPGW-prb-2007,bobbert-vertex_GW-prb-1994,kutepov-vertex_GW-prb-2016}. As a result, $GW\Gamma$ often produces results similar to $G_0W_0$ \cite{delsole-GWGamma-prb-1994,mahan-GWGamma-prl-1989}. Even looping on $G$ without vertex corrections can sometimes worsen results because the $Z$-factor renormalization underestimates the self-energy correction \cite{rodl-vertex_GW-prb-2015,kotani-QPGW-prb-2007}. The off-diagonal part of the self-energy is typically less important, as the converged wavefunctions are often close to the HF wavefunctions. Ultimately, the most reliable criterion for choosing a scheme is agreement with experimental data. Comparison of indirect and direct gaps and bandwidths from $GW$ calculations with experiments is standard, though such data are often lacking for moir\'e systems. 

Moreover, most sophisticated $GW$ schemes were developed for insulators or semiconductors with gaps on the order of hundreds of meV, whereas correlated gaps in moir\'e systems are typically only tens of meV with comparable bandwidths. Previous studies indicate that gaps obtained by different $GW$ schemes differ by at most 10\% \cite{hybertsen-GW0-prb-1986,hybertsen-GW-prl-1985}, which corresponds to 1\;meV in our case. Our recent calculations for R5G \cite{lu-MBPTmoire-arxiv-2025} show consistent results across schemes without significant discrepancies \cite{rodl-vertex_GW-prb-2015,bruneval-compareGW-prb-2006}. Therefore, in our theory, we adopt the EV-$GW$ scheme, calculating $GW$ quasiparticle bands separately, as we only consider the diagonal part of the self-energy.

\subsubsection{Multiple plasmon pole approximation}

The most numerically demanding part of the EV-$GW$ method is the calculation of the polarizability, which involves a double summation over bands (see Eq.~\eqref{eq:chi0}) at multiple frequencies to compute the self-energy using Eq.~\eqref{eq:selfenergy_corr}. Due to the slow convergence with respect to band index, as encountered in the RPA correlation energy, no cutoff can be imposed on the summation. Therefore, all bands within the plane-wave cutoff must be included, highlighting once again the necessity of all-band HF calculations. A naive evaluation of the self-energy by computing $\chi_0$ at many frequencies to ensure convergence would be extremely computationally demanding, especially in self-consistent EV-$GW$ loops where quasiparticle energies must be updated both in $G$ and in $\chi$ (hidden in $W$). This motivates the use of an approximation that simplifies the numerical evaluation of the self-energy: the multiple plasmon-pole approximation (MPA) \cite{mpa-prb21,mpa-prb23}.

The MPA, first proposed by Leon et al. \cite{mpa-prb21,mpa-prb23}, is an improvement over the single plasmon-pole model commonly used to approximate the dielectric function \cite{louie-ppa-prb86,hybertsen-ppa-prb89,needs-ppa-prl89,horsch-ppa-prb88,engel-ppa-prb93}. The single plasmon-pole model is based on the observation that the pole of the self-energy coincides with that of the polarizability, which near the pole has a functional form $\sim 1/(\nu^2-\Omega^2)$. The self-energy can then be approximately evaluated using the residue of this single pole. In contrast, the MPA assumes the existence of multiple particle-hole collective excitations and fits the dielectric function using these multiple modes. By capturing both discrete collective modes and the continuum spectra, the MPA provides a more accurate description of the system's collective excitations. This approach not only improves the precision compared to the single plasmon-pole model but also greatly enhances computational efficiency by reducing the complexity of frequency integration.

The full derivation of the MPA is given in literature \cite{mpa-prb21,mpa-prb23}. Our theory exactly follows these works \cite{guo-AHC-arxiv-2025}. Here, we give directly the functional form of $\Sigma_c$
\begin{equation}
	\epsilon^{-1,MPA}_{\Q\Q'}(\btq,\nu)-\delta_{\Q\Q'}
	=\sum^{N_p}_{l}\frac{2R_{l,\Q\Q'}(\btq)\Omega_{l,\Q\Q'}(\btq)}{\nu^2-\Omega^2_{l,\Q\Q'}(\btq)},
    \label{eq:MPA}
\end{equation}
where $N_p$ represents the number of plasmon poles, while $R$ and $\Omega$ are parameters to be determined by fitting. Specifically, $R_{l,\Q\Q'}(\btq)$ denotes the residue of the $l$-th plasmon mode, and $\Omega_{l,\Q\Q'}(\btq)$ represents its frequency. To solve for these unknown parameters, we require the values of the dielectric function at $2N_p$ different frequencies. The choice of these frequencies is given in literature \cite{mpa-prb21,mpa-prb23}.

The advantage of the MPA is that it can simultaneously describe both discrete plasmon modes and the continuous spectrum, eliminating the need for complicated frequency integrations and greatly accelerating the calculation. Moreover, the accuracy of the approximation can be systematically controlled by adjusting the number of poles, allowing for a balance between computational efficiency and precision. This is particularly useful for spontaneous-symmetry-breaking states such as the Wigner crystal, which feature multiple collective modes, including acoustic and optical quantum phonons \cite{macdonald-wc-prb91}.

\begin{figure*}
    \centering
    \includegraphics[width=0.95\textwidth]{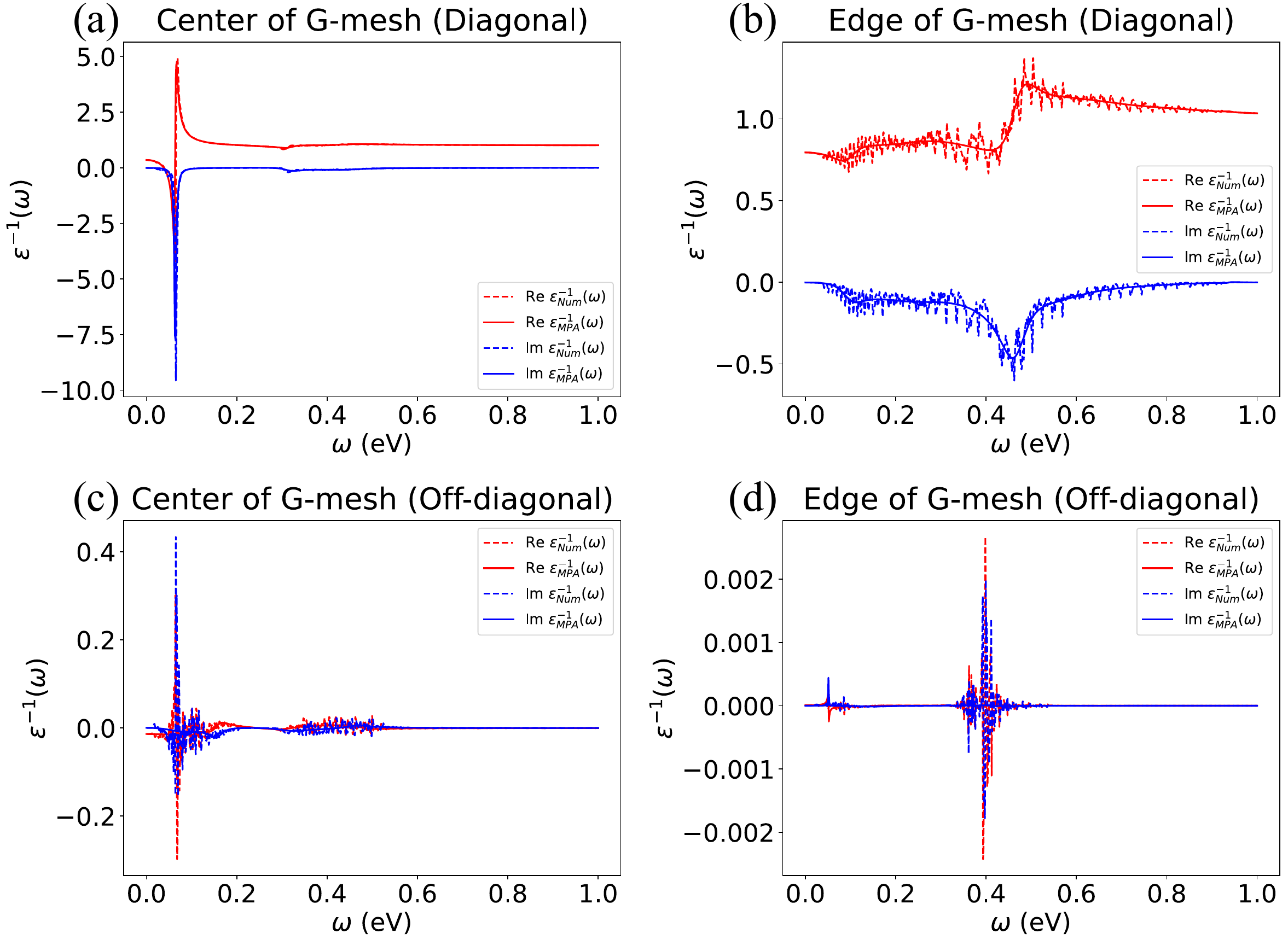}
    \caption{Comparison between the numerically calculated inverse dielectric function (dashed lines) and the MPA-fitted inverse dielectric function (solid lines). The real part is shown in red, and the imaginary part is shown in blue. (a) Small $\btq$ with $\G = 0$ near the $\Gamma$ point, showing a prominent plasmon mode near $q = 0$. (b) Non-zero $\G$ away from the $\Gamma$ point, entering the continuum spectrum region. (c) and (d) Off-diagonal elements near and away from the $\Gamma$ point. Figure extracted Zhongqing Guo and Jianpeng Liu, arXiv:2409.14658, 2024 \cite{guo-AHC-arxiv-2025}, with formal permission from
    the corresponding author.}
    \label{fig:MPAfit}
\end{figure*}

Fig.~\ref{fig:MPAfit} illustrates a comparison between the numerically calculated inverse dielectric function (dashed lines) and the MPA-fitted inverse dielectric function (solid lines). The red and blue lines correspond to the real and imaginary parts, respectively. In Fig.~\ref{fig:MPAfit}(a), for small $\btq$ with $\G = 0$ near the $\Gamma$ point, a clear plasmon mode is visible, characterized by a peak in the imaginary part and a singularity in the real part near $q = 0$. The MPA accurately captures this plasmon mode while also reproducing the continuum part of the dielectric function at higher frequencies. In Fig.~\ref{fig:MPAfit}(b), for $\G \neq 0$ away from the $\Gamma$ point, the system enters the continuum spectrum region of the inverse dielectric function. The MPA successfully represents the continuous spectrum using multiple plasmons, a capability beyond the single plasmon-pole model. Additionally, oscillations in the numerically computed inverse dielectric function, arising from $\bk$-mesh discretization or finite-size effects, are smoothed out by the MPA. These small discrepancies have negligible impact on the final self-energy integral. Figs.~\ref{fig:MPAfit}(c) and (d) show the off-diagonal elements of the inverse dielectric function, both near and far from the $\Gamma$ point. Although these elements are relatively small, the MPA still provides an accurate description.

Using the MPA, we can compute $\Sigma_c$ more efficiently without losing too much accuracy. Within the MPA, $\Sigma_c$ is given by:
\begin{equation}
	\begin{aligned} 
		&\Sigma_{c}(\btk,\omega)_{nn}= \frac{1}{N_s\Omega_0}\sum_{m,\btq}\sum_{\Q,\Q'} \Lambda^*_{mn}(\btq,\Q) \Lambda_{mn}(\btq,\Q') \\
		&\times \sum^{N_p}_{l}\frac{V(\btq+\Q)R_{l,\Q \Q'}(\btq)}{\omega-\varepsilon_{m\btk+\btq}+i\delta_{m\btk+\btq}+\Omega_{l,\Q \Q'}(\btq)(2f_{m\btk+\btq}-1)},
	\end{aligned}
	\label{eq:self-energy_mpa}
\end{equation}
where $f_{m\btk+\btq}$ represents the Fermi-Dirac distribution function. The derivative of the correlation self-energy with respect to $\omega$ is required to determine the quasiparticle weight $Z_{n\btk}$:
\begin{equation}
	\begin{aligned}
		&\frac{\partial \Sigma_{c}(\btk,\omega)_{nn}}{\partial \omega}=\frac{-1}{N\Omega_0}\sum_{m,\btq}\sum_{\Q,\Q'} \Lambda^*_{mn}(\btq,\Q) \Lambda_{mn}(\btq,\Q')\\
		&\times \sum^{N_p}_{l}\frac{V(\btq+\Q)R_{l,\Q \Q'}(\btq)}{\left[ \omega-\varepsilon_{m\btk+\btq}+i\delta_{m\btk+\btq}+\Omega_{l,\Q \Q'}(\btq)(2f_{m\btk+\btq}-1)\right] ^2}.
	\end{aligned}
	\label{eq:self-energy-diff_mpa}
\end{equation}

\subsection{Example: R5G-hBN}
\begin{figure}
    \centering
    \includegraphics[width=0.45\textwidth]{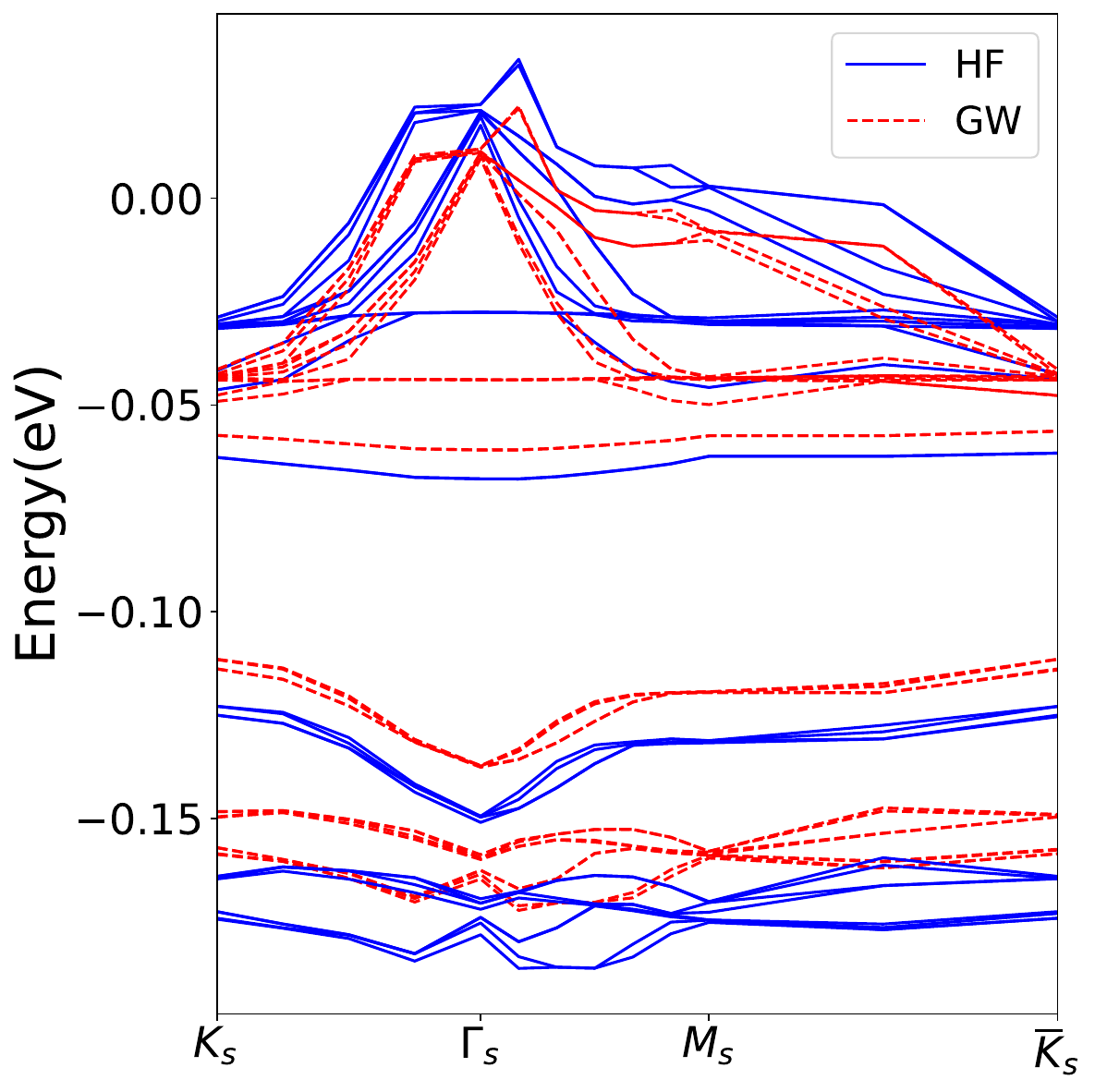}
    \caption{Comparison between the HF single particle bands calculated by all-band HF method and the $GW$ quasiparticle bands for hBN-R5G under electric field $D=0.7$\;V/nm. The quasiparticle weight for this particular case is above 0.9.}
    \label{fig:GWband}
\end{figure}

Taking R5G-hBN as an illustrative example, Fig.~\ref{fig:GWband} shows how the $GW$ approximation, evaluated using the MPA, affects the single-particle spectra. We observe that the valence bands shift upward in energy, while the conduction bands shift downward, with this shift being nearly uniform across $k$-space. This effect, commonly referred to as the scissor operator \cite{godby-GW-prb-1988}, has been widely observed in $GW$ calculations even before the advent of moir\'e superlattices. As a result, the indirect and direct gaps are reduced by approximately 60\%, and the bandwidth decreases by about 30\%.

More importantly, the quasiparticle weight is found to be around 0.9, comparable to silicon where the quasiparticle weight is approximately 0.8 according to DFT calculations \cite{hybertsen-GW0-prb-1986}. This indicates that HF approximations already provide a reasonably accurate description of moir\'e systems at integer fillings. This is consistent with what have been found using exact diagonalization \cite{yu-R5GED-prb-2025}, which suffers from the finite size effects.  

\subsection{Alliance of $GW$ and RPA correlation energy}
By combining the $GW$ approximation with the RPA correlation energy, one can, in principle, achieve a more comprehensive description of the total energy. The $GW$ approximation improves the single-particle energy spectrum, yielding more accurate values for $\chi^0(i\omega)$ in Eq.~\eqref{eq:EcRPA}, so that the RPA, which incorporates dynamical charge fluctuations, can provide a more reliable estimate of the correlation energy. This combined approach is particularly important for capturing the delicate balance between the Fermi liquid state and competing spontaneous symmetry-breaking phases, such as the Wigner crystal, especially in low-carrier-density interacting two-dimensional systems \cite{guo-AHC-arxiv-2025}. However, convergence of the RPA correlation energy with respect to the $GW$ quasiparticle band cutoff can only be achieved when nearly all bands are included, which is computationally demanding. At the current state of the art, without further approximations, $GW$+RPA correlation energy calculations are feasible only for a few specific cases and remain challenging for large-scale calculations, such as generating phase diagrams of moir\'e systems.

Thanks to rapid experimental progress, the HF+$GW$+RPA workflow can now be directly validated in TBG. A key enabling development is the recent invention of the quantum twisting microscope \cite{birkbeck-QTM_TBG-nature-2025}. Intuitively, this technique, which is based on scanning tunneling microscopy, provide momentum-resolved spectral information similar to angle-resolved photoemission spectroscopy. Most saliently, this technique allows the moiré bands in TBG to be resolved across the moiré Brillouin zone with meV-scale energy resolution \cite{xiao-QTM_TBG-arxiv-2025}. In our recent work, we applied the HF+$GW$+RPA framework to TBG and obtained quantitative agreement with experimental observations, including band gaps, bandwidths, and band shapes \cite{lu-MBPTmoire-arxiv-2025}.

Nevertheless, many-body perturbation theory primarily provides a more quantitatively accurate description of moir\'e systems. The phases it predicts remain those obtained from HF calculations. The improvement lies in producing more realistic quasiparticle bands and total energies. Since the method treats correlations perturbatively, it cannot capture strongly correlated phases, where correlations dominate over exchange interactions—such as FCI states, which exemplify the interplay between strong correlations and topology. This limitation motivates the use of exact diagonalization, as discussed in the following section.

\section{Exact diagonalization}
Exact diagonalization (ED) is a method to obtain the many-body ground state by explicitly constructing and solving the full interacting Hamiltonian in the many-body Hilbert space. It provides the most direct and unbiased treatment of quantum many-body problems, without relying on approximations for quartic interacting terms. The main limitation is the rapid growth of the Hilbert space with system size, which restricts the number of electrons that can be treated. Although one can reduce the Hilbert space by exploiting symmetries of the problem, such as particle number conservation and translational symmetry, the dimension of the Hamiltonian matrix still grows exponentially—a phenomenon often referred to as the ``exponential wall''. For example, for 12 spinless electrons distributed among 36 sites, the Hilbert space dimension exceeds one billion, approaching the limit of what modern high-performance computing can handle. Nevertheless, ED remains one of the most reliable techniques for studying gapped strongly correlated phases such as FCIs, which cannot be treated by any of the mean-field-based methods discussed previously.

In this section, we focus on subtleties of implementing ED specifically for FCI states observed experimentally in moir\'e systems. First, we describe how to identify a relevant subspace of the full Hilbert space to alleviate computational costs, including the typical arguments that justify such simplifications. Next, we discuss techniques to address the double-counting problem that arises if HF states are used as building blocks for ED. FCI states have been unambiguously observed in hBN-R$n$G heterostructures \cite{fqah-julong-nature24,fqah-julong-nature25,fqah-luxb-natmat25} and tTMD systems \cite{fqah-optics-xu-nature23,fqah-mak-nature23,fqah-nature23,fqah-prx23}. In hBN-R$n$G, fractionally quantized Hall resistance appears under a strong vertical electric field at fractional moiré filling $\nu<1$ of the conduction band, whereas in tTMD, a fractional quantum anomalous Hall effect is observed in the valence moir\'e band at fractional hole filling $|\nu|<1$. Here, we limit our discussion to the case $|\nu|<1$. For general numerical techniques, such as binary encoding of quantum states, the Lanczos algorithm for large sparse matrices, and parallelization strategies, we refer the reader to specialized literature \cite{fehskhe-book}.

\subsection{Reduce the dimension of Hilbert space}
Due to numerical limitations, the ideal scenario for studying FCI states using ED is to identify a single, isolated moir\'e band that can be partially filled. Achieving this in moir\'e systems requires careful consideration because of the valley degeneracy protected by time-reversal symmetry. The most striking example is TBG, where eight flat bands—arising from spin, valley, and orbital degrees of freedom—collapse into a tiny energy window around the CNP in an inseparable manner. Nevertheless, it can be argued that strong exchange interactions favor spontaneous polarization into a single flat band with a specific isospin flavor, meaning electrons tend to align along the same (iso)spin direction. Note that this argument does not strictly hold even at integer fillings, where intervalley-coherent states can emerge as HF ground states. However, experimental observations indicate that FCI states appear only within parameter regimes where integer Chern insulator states are also realized, supporting the applicability of this assumption for tTMD and hBN-R$n$G heterostructures.

Once such an isolated band is identified, Coulomb interactions are projected onto this subspace. Ideally, the energy gap separating this band from others should exceed the Coulomb interaction strength (tens of meV). In practice, however, the gap is often comparable to the interaction energy. For the moment, we assume that this does not qualitatively affect the results and leave a detailed discussion of multiband effects to the last part of the section.

As an example, consider $\nu = 1/3$ filling with 12 electrons distributed among 36 $\bk$ points in the moir\'e Brillouin zone. The dimension of the ED Hamiltonian can be further reduced by classifying states according to the total momentum modulo the moir\'e reciprocal lattice vector, which decreases the Hamiltonian size by a factor equal to the number of sites. Unlike the quantum Hall problem on a torus, one cannot exploit additional many-body translation symmetries in lattice systems, which is an important distinction. Direct calculations show that the final Hamiltonian matrix has a dimension of approximately 35 million, which is now manageable using state-of-the-art ED techniques for phase diagram computations.

\subsection{Choice of momentum mesh}

The chosen $\bk$-mesh should be as evenly distributed as possible along the $x$ and $y$ directions, i.e., with an aspect ratio close to unity. This is important because finite-size effects can significantly influence the results, as well-known in previous lattice-model investigations of FCIs \cite{regnault-fci-prx11,lauchli-fci-prl-2013,repellin-prb14}. It is also preferable to select $\bk$ points that cover the high-symmetry points of the Brillouin zone, which is crucial for capturing commensurate charge density wave states—the main competing phases to FCIs. For example, in a triangular lattice, as is the case for TMD and graphene, the first Brillouin zone is a lozenge with one angle equal to $\pi/3$. Therefore, it is advantageous to choose the number of $\bk$ points as a multiple of six, making a $6 \times 6$ $\bk$-mesh an ideal choice.

In practice, numerical limitations may prevent the use of a perfect $6 \times 6$ $\bk$-mesh. For instance, for a fractional filling of $1/2$, the Hilbert space of 18 spinless electrons over 36 sites—even after exploiting lattice translation symmetry—exceeds 25 billion states. Additionally, calculations with varying numbers of $\bk$ points are needed to perform finite-size scaling and extrapolation of many-body properties.

To balance the number of sites with a desirable aspect ratio, one can employ tilted boundary conditions to construct the $\bk$-mesh. This approach was first proposed by Laulich et al. for square lattices \cite{lauchli-fci-prl-2013} and later generalized by Repellin et al. to arbitrary lattices \cite{repellin-z2fci-prb-2014}. The general idea is to first construct a supercell from the original Bravais lattice vectors, which in our case are the moiré superlattice vectors ${\mathbf{R}_1, \mathbf{R}_2}$, and then define the $\bk$ points accordingly. Then, the supercell lattice vectors can be generically written as 
\begin{equation} 
    \mathbf{L}_1 = n_{11}  \mathbf{R}_1 + n_{12}  \mathbf{R}_2 \quad , \quad \mathbf{L}_2 = n_{21}  \mathbf{R}_1 + n_{22}  \mathbf{R}_2
    \label{eq:supercell}
\end{equation} 
where $n_{ij}$ with $i,j=1,2$ are integers. The number of sites $N_s$ can be then set as $\left|\det (n_{ij}) \right|$. The reciprocal lattice vectors $\mathbf{G}_{1,2}$ of $\mathbf{L}_{1,2}$ already defines a $\bk$-mesh, but the aspect ratio can be further improved by searching the possibility of $\mathbf{G}_{1,2}'$ such that
\begin{equation} 
    \mathbf{G}_1' = m_{11}  \mathbf{G}_1 + m_{12}  \mathbf{G}_2 \quad , \quad \mathbf{G}_2' = m_{21}  \mathbf{G}_1 + m_{22}  \mathbf{G}_2
    \label{eq:supercellGprime}
\end{equation} 
where $m_{ij}$ with $i,j=1,2$ are integers and $\det (m_{ij})=1$. The $\bk$-mesh is defined by
\begin{equation}
    \bk = r_1 \mathbf{G}_1' + r_2 \mathbf{G}_2'
    \label{eq:kmesh}
\end{equation}
where integers $r_{1,2} \in [0,N_{1,2}-1]$ with $N_s=N_1 \times N_2$. The final choice of $\bk$ mesh is built on the one that have an aspect ratio closest to unity among all the possible $m_{ij}$. In practice, it suffices to impose $m_{ii}=1$ and $m_{12}=0$ and then find the best $m_{22}$.

\subsection{Double counting of interactions}
To study FCI states in R5G-hBN heterostructures, we encounter another challenge: even when considering only the moir\'e bands for a single valley and single spin, the first conduction band is entangled with higher conduction bands, as shown in Fig.~\ref{fig:HF_R5G_hBN}(b). A Chern-number-1 isolated conduction band emerges only after HF calculations at $\nu=1$, where Fock exchange interactions are responsible for opening a gap between the first conduction band and the higher conduction bands. This isolated HF band with Chern number 1 is commonly considered the precursor of FCI states.

However, one cannot perform ED by simply projecting onto the lowest HF conduction band. This is because the HF calculations already include part of the interaction effects at the mean-field level, which in principle should not be double-counted in the ED starting from the HF states at $\nu=1$. To address this issue, three treatments have been proposed in literatures, to the best of our knowledge. One way is to subtract the HF contribution of interaction $\hat{V}_{H/F}$ from the full interaction $\hat{V}_c$ before projection \cite{dong-R5Gfci-prl-2024}, namely
\begin{equation}
    \hat{H}_{\rm ED} \approx \sum_\bk \varepsilon^{\rm HF}_{\rm{1CB},\bk} \hcd_{\rm 1CB} (\bk) \hc_{\rm 1CB} (\bk)  +  \hat{P} ( \hat{V}_c - \hat{V}_{H} - \hat{V}_{F}) \hat{P}
    \label{eq:EDsubHF}
\end{equation}
where the projection operator $\hat{P}$ to the first conduction band reads $\sum_\bk \ket{\psi^{\rm HF}_{\rm 1CB}(\bk)}\bra{\psi^{\rm HF}_{\rm 1CB} (\bk)}$. 

Another way to justify performing ED on the projection to the first HF conduction band is to adopt a hole-picture perspective instead of electrons \cite{zhou-R5Gfci-prl-2024,guo-HFfci-prb-2024}. The key observation is that the Chern-number-1 HF conduction band is entirely empty of holes, so a fractional electron filling $\nu$ corresponds to a fractional hole filling $\bar{\nu} = 1 - \nu$. When projecting onto this empty hole band, we neglect the presence of all other holes, since the band is well isolated from the rest. Consequently, there is no double-counting issue when partially filling this isolated band with holes, because the starting point is an empty band. The Hamiltonian to solve using ED becomes
\begin{equation}
    \hat{H}_{\rm ED} \approx \sum_\bk -\varepsilon^{\rm HF}_{\rm{1CB},\bk} \hcd_{\rm 1CB} (\bk) \hc_{\rm 1CB} (\bk)  +  \hat{P} ( \hat{V}_c) \hat{P} \; .
    \label{eq:EDhole}
\end{equation}
However, this argument neglects particle-hole symmetry-breaking terms present in the continuum model as well as in the Coulomb interactions due to form factors. Both of the methods mentioned above implicitly assume that the single-particle orbitals at fractional filling remain close to the HF orbitals at $\nu=1$.

The third approach for performing ED is to carry out constrained HF calculations directly at the fractional filling by assuming that the first conduction band is uniformly occupied, i.e., $\langle n(\mathbf{k}) \rangle = \nu$, as expected for liquid states such as the FCI \cite{huang-fqahdoublecounting-prb-2024}. The ED is then performed on these specially prepared HF orbitals. This procedure is equivalent to treating interband interactions between the valence and conduction bands at the HF level, while treating intraband interactions within the lowest conduction band at the ED level.

Fortunately, although these three treatments appear to differ significantly, numerical calculations show that all three approaches lead to qualitatively similar results: the FCI states emerge as the ground state in the experimentally relevant parameter regime.

\subsection{Determination of many-body ground states}
Solving the projected many-body Hamiltonian yields the energy spectrum of many-body states, which can be classified according to their total momentum sectors. A key feature of an FCI is the quasi-degeneracy of the ground-state manifold, which is separated from higher-energy states by a finite gap. The degeneracy of the ground states depends on the filling factor. Here, we focus on Laughlin-type FCIs. For instance, for a $\nu=1/3$ Laughlin-type FCI, the ground-state manifold exhibits a three-fold degeneracy.

However, both FCI and charge density wave (CDW) states can exhibit (nearly) degenerate ground states and finite energy gaps. Several criteria can be used to distinguish them. First, the total momenta of the quasi-degenerate states must satisfy the ``generalized Pauli principle'' \cite{regnault-fci-prx11}, which mimics the rules obeyed by Laughlin states in the quantum Hall problem on a disk \cite{bernevig-jackPauli-prl-2008}. Second, under the adiabatic insertion of a magnetic flux $\phi$, the three degenerate ground states of an FCI evolve into each other and return to the original configuration when $\phi = 3\phi_0$ ($\phi_0 = h/e$ is the flux quantum), a behavior absent in CDW states \cite{regnault-fci-prx11}. In the context of moir\'e systems, flux insertion corresponds to a shift in the crystalline momentum along one direction, for example $\delta k_1 = \phi/\phi_0$, where $k_1$ is the reduced crystalline momentum along the first reciprocal vector direction.

Another distinguishing feature is the charge distribution in real space. An FCI is an incompressible liquid, and thus the momentum-space occupation is nearly uniform, $\langle n(\mathbf{k}) \rangle = \nu$. In contrast, a CDW exhibits non-uniform momentum-space occupation, where sites with $\langle n(\mathbf{k}) \rangle = 1$ define the periodicity of the CDW in real space \cite{regnault-fci-prx11}. Similarly, the structure factor $S(\mathbf{q})$, defined as the Fourier transform of the pair correlation function, provides a clear distinction \cite{dong-cflTMD-prl-2023,wilhelm-fcicdwTBG-prb-2021}: an FCI exhibits a single peak at $\mathbf{q} = 0$, whereas a CDW shows additional Bragg peaks at $\mathbf{q} \neq 0$, often at high-symmetry points of the Brillouin zone.

It is important to note that the projection procedure restricts the Hilbert space for ED to many-body states constructed from the single-particle orbitals of the Chern-number-1 first HF conduction band. However, there is no physical guarantee that many-body states built from single-particle orbitals of HF metastable states at $\nu=1$ (usually Chern-number-0 gapped states) always have higher energy than those from Chern-number-1 states. Therefore, to identify the genuine many-body ground state, one should compare the energies of the ground states obtained from these two different projected Hilbert spaces.

\subsection{Multiband effects}
The discussion of interband coupling in tTMD and hBN-R$n$G within the framework of ED is currently limited by computational capabilities. There have been some attempts to analyze how single-band results should be modified when multiband effects are included. One possible approach is to slightly expand the Hilbert space, allowing a few particles to occupy the first and second higher bands \cite{yu-R5GED-prb-2025}. From this, one can self-consistently determine a set of optimized single-particle orbitals by maximizing their occupation numbers, such that the results of the multiband ED can be reasonably reproduced by projecting the Hamiltonian onto this optimized basis \cite{li-multibandED-prb-2025}.

Studies have found that the inclusion of multiband effects tends to reduce the gap separating the ground-state manifold from higher-energy states \cite{yu-R5GED-prb-2025,li-multibandED-prb-2025}. Similar conclusions have also been reported in recent works employing deep-learning-based quantum Monte Carlo methods \cite{li-deeplearningFCITMD-arxiv-2025,qian-deeplearningFQH-prl-2025}. It is worth noting that the energy separation between flat bands and remote bands is often comparable to the strength of Coulomb interactions, such that interband mixing can be significant and non-negligible. In some cases \cite{yu-R5GED-prb-2025}, including multiple bands can collapse the FCI gap through fluctuation-driven mixing, ultimately destabilizing the FCI phases. This suggests that the current single-band ED framework likely overestimates the stability of FCI phases. However, given the experimental stability of FCI phases, it is natural to conjecture that some mechanism embedded in the interband coupling contributes to stabilizing these phases. For example, the many-body wavefunction may involve single-particle components from high-energy bands to facilitate the emergence of FCI, as has already been demonstrated in the context of the quantum Hall problem \cite{qian-deeplearningFQH-prl-2025}. Disorder may also play important role, which may induce a thermal crossover to fractional Chern insulator state \cite{thermal-fci-sarma-prb24,thermal-fci-li-arxiv24}. Therefore, further investigation is certainly required before reaching a definitive conclusion about multiband effects, and a comprehensive treatment lies beyond the scope of the present manuscript.

\section{Discussions}
While superconductivity has been observed in various moiré systems \cite{cao-nature18-supercond,chen-hbn-trilayer-nature19,park2021tunable,hao-SCtmg-science-2021,zhang-SCtmg-science-2022,park-SCtmg-natmat-2022,dean-wse2-nature25,mak-wse2-nature25,xu-SCmote2-arxiv-2025}, the most prominent example is TBG, which has been extensively studied due to the availability of a wide range of experimental results across different twist angles and carrier dopings \cite{cao-nature18-supercond,cao-nature18-mott,dean-tbg-science19,efetov-nature19,efetov-nature20,young-tbg-np20}. This is likely why TBG has become the central focus of theoretical studies on superconductivity in moiré materials. Below, we also use TBG as a representative example to briefly discuss recent advances in computational methods in this field.

The key element of superconductivity is the presence of attractive interaction channels that bind two or more electrons into pairs, which then condense leading to superfluidity. Several mechanisms have been proposed for such attractive channels in TBG, which can broadly be categorized into two classes: those mediated by $e$-$e$ interactions \cite{balents-npreview20,dodaro-SCtbg-prb-2018,xu-balents-prl18,liu-prl18,guo-SCtbg-prb-2018,lothman-tbgSC-comphys-2022,guinea-tbg-pnas18,you-tbgSC-npjQM-2019,gonzalez-tbgSC-prl-2019,wang-tbgSC-prb-2021,khalaf-skyrmionTBG-sciadv-2021,chatterjee-TBGskyrmionSC-prb-2022,yu-tbgSC-prb-2023,islam-tbgSC-prb-2023} and those mediated by electron-phonon coupling \cite{wu-prl18,lian-tbg-prl19,peltonen-SCtbg-prb-2018,blason-tbgSC-prb-2022,christos-tbgSC-natcom-2023,liu-tbg_epc-prb-2024,wang-tbg_epc-prb-2025,wang-tbgSC-prl-2024}. In the former case, the pairing mechanism is attributed to an effective attractive channel arising from purely electronic interactions. Early studies used effective lattice Hubbard models \cite{dodaro-SCtbg-prb-2018,xu-balents-prl18,liu-prl18,guo-SCtbg-prb-2018}, where the non-interacting part mimics the flat-band dispersion of TBG, and extended the Hubbard model's knowledge to TBG using Bogoliubov-de Gennes (BdG) mean-field approach, RPA, quantum Monte Carlo, etc. However, this approach overlooks the topological nature of TBG's flat-band wavefunctions, which is a critical issue since the quantum geometric contribution to the superfluid weight \cite{torma-QGandSC-natrevphys-2022} is comparable to the conventional contribution, as both theory \cite{hu-tbg-prl19,torma-prb20} and experiments \cite{tian-TBGQGSC-nature-2023} indicate. Some groups \cite{lothman-tbgSC-comphys-2022} have attempted full-scale calculations based on atomic tight-binding models, but these are currently only feasible down to twist angles of about 1.2$^\circ$. Consequently, theoretical studies of superconductivity in TBG are still more tractable when based on continuum models.

Efforts have been made to develop BdG mean-field and RPA approaches within the continuum description to account for superconducting instabilities in TBG, through mechanisms such as Kohn-Luttinger-type mechanism or flavor fluctuations \cite{guinea-tbg-pnas18,you-tbgSC-npjQM-2019,gonzalez-tbgSC-prl-2019,wang-tbgSC-prb-2021,yu-tbgSC-prb-2023,islam-tbgSC-prb-2023}, and to propose effective low-energy field theories of Skyrmion 2$e$ condensation \cite{khalaf-skyrmionTBG-sciadv-2021}, supported by density matrix renormalization group numerics \cite{chatterjee-TBGskyrmionSC-prb-2022}. Experimental evidence of charged Skyrmion excitations with respect to correlated insulator at filling 2 is also reported in TBG \cite{yu-TBGskyrmion-natcom-2023}. Nevertheless, many experimental observations also seem to support a phonon-mediated pairing mechanism. The unusual aspects of superconductivity in TBG are likely related to its flat band nature \cite{peltonen-SCtbg-prb-2018}. For instance, screened Coulomb interactions tend to suppress the gap associated with correlated insulating phases but, conversely, enhance superconductivity \cite{efetov-nature20,young-tbg-np20,li-tbg-science21}. Recent experiments also provide evidence of strong electron-phonon coupling in superconducting devices of magic-angle TBG \cite{chen-replica-nature24}. This highlights the importance of deriving the full electron-phonon coupling vertex from microscopic tight-binding models \cite{liu-tbg_epc-prb-2024,wang-tbg_epc-prb-2025} and consistently implementing it in continuum descriptions—an area that remains underdeveloped. Moreover, electron-electron interactions and electron-phonon couplings, both of which seem to play important roles in the physics of magic-angle TBG, need to be treated on equal footing, to develop a faithful and ``first principles'' theory of superconductivity. Specifically, such an approach requires careful treatment of dynamic screening in both electron-electron and electron-phonon interaction channels.

Alternative approaches, such as topological heavy-fermion models \cite{song-heavyfermion-prl22}, have also been proposed. By projecting the electron-phonon coupling onto the topological heavy-fermion basis \cite{song-heavyfermion-prl22,wang-tbgSC-prl-2024,wang-tbg_epc-prb-2025}, one can reconcile the coexistence of Kondo physics and superconductivity, both observed in TBG \cite{saito-pomerchunkTBG-nature-2021,rozen-pomerchunkTBG-nature-2021}. This framework particularly explains how a relatively weak phonon-mediated attraction ($\sim 1$ meV) can overcome the seemingly prohibitive Coulomb repulsion (tens of meV) \cite{wang-tbgSC-prl-2024}.

In contrast to twisted graphene systems, robust evidence of superconductivity in twisted TMD systems has only recently been observed in twisted WSe$_2$ \cite{mak-wse2-nature25,dean-wse2-nature25} and twisted MoTe$_2$ \cite{xu-SCmote2-arxiv-2025}. The superconducting mechanisms in these two systems are likely different. In twisted WSe$_2$, superconductivity has been observed across various devices with twist angles ranging from 3.5$\circ$ to 5$^\circ$, suggesting that superconductivity is not highly sensitive to twist angle, and thus, is less likely to be associated with the formation of flat bands. The reduction in the bandwidth of the low-energy bands from 5$^\circ$ to 3.5$^\circ$ indicates that superconductivity in twisted WSe$_2$ is likely driven by a weak-coupling pairing mechanism at 5$^\circ$, while it may be governed by a strong-coupling mechanism at 3.5$^\circ$. Current theoretical studies are also based on continuum models \cite{qin-SCtmd-arxiv-2024,guerci-SCtmd-arxiv-2024,zhu-SCtmd-prb-2025,christos-SCtmd-prl-2025} and effective lattice Hubbard models \cite{tuo-SCtmd-natcom-2025,fischer-SCtmd-arxiv-2024,wu-SCtmd-prl-2023,xie-SCtmd-prl-2025,kim-SCtmd-natcom-2025}. Unlike TBG, the low-energy bands of twisted WSe$_2$ can be Wannierized from the continuum model, which includes only a few orbitals. Studies based solely on continuum models focus on Coulomb $e$-$e$ interaction-driven mechanisms using the BdG mean-field approach combined with RPA-screened interactions, and suggest that intervalley pairing is responsible for superconductivity in twisted WSe$_2$ \cite{qin-SCtmd-arxiv-2024,guerci-SCtmd-arxiv-2024,zhu-SCtmd-prb-2025}. Those using extended Hubbard models suggest other pairing mechanisms that lead to topological superconductivity, using mean-field theory combined with RPA, RG, DMFT, etc. \cite{tuo-SCtmd-natcom-2025,fischer-SCtmd-arxiv-2024,wu-SCtmd-prl-2023,xie-SCtmd-prl-2025,kim-SCtmd-natcom-2025}.

The situation in twisted MoTe$_2$ is somewhat different, as the superconducting phase is surrounded by a ferromagnetic metal and adjacent to FCI states and reentrant Chern insulator states \cite{xu-SCmote2-arxiv-2025}. This suggests that pairing may occur within a valley-polarized, ferromagnetic state, giving rise to chiral superconductivity. So far, few theoretical works have addressed this problem, but some propose a Kohn-Luttinger mechanism that leads to chiral superconductivity, using BdG mean-field theory combined with RPA screening \cite{xu-SCtmd-arxiv-2025}. Regardless, more experimental evidence and theoretical calculations are needed to fully clarify the mechanism of superconductivity in these two twisted TMD systems. It is worth noting that evidence of chiral superconductivity is also reported in moir\'eless graphene systems including both tetralayer and pentalayer rhombohedral graphene under slight electron dopings \cite{Han-nature2025}, without alignment with the hBN substrate. The mechanism of the possible chiral superconductivity emerging in these systems remains an open question, and is an active research area \cite{wen-chiral-prb25,xu-SCtmd-arxiv-2025,eslam-anyon-arxiv25,ashvin-anyon-arxiv25}.

From a methodological perspective, the theoretical framework presented in this work should also be inspiring for future studies of superconductivity. Regardless of the specific superconducting mechanism, a reliable continuum model is essential. Moreover, dynamic screening, which plays a crucial role in determining the attractive channel, can be addressed by extending all-band HF approximations and $GW$ approximations. Exact diagonalization should be useful for explaining or at least providing insight into superconductivity near integer filling or commensurate fractional fillings, within a unified framework.

In conclusion, this work presents a comprehensive theoretical workflow for studying moir\'e systems (shown by Fig.~\ref{fig:workflow}), combining continuum modeling, Hartree-Fock approximations, many-body perturbation theory, and exact diagonalization. By carefully implementing these numerical techniques and considering the subtleties specific to moir\'e systems, we provide a framework that connects theoretical predictions with experimental observations. While each method has its own limitations, their combined application allows for a systematic investigation of correlated and topological phases, including correlated insulators and FCIs. 

Looking forward, further development of these methods—particularly the incorporation of multiband effects in exact diagonalization and more realistic dynamical screening in many-body perturbation theory—will be essential for fully capturing the rich physics of moir\'e systems. Moreover, incorporing the phonon degrees of freedom and electron-phonon couplings in a unified framework may also be crucial, especially to shed light on the microscopic mechanism of superconductivity. We hope this work not only clarifies the technical aspects of the commonly used theoretical approaches but also serves as a practical guide for researchers, bridging the gap between theory and experiment, and inspiring further theoretical advancements in this rapidly evolving field.

\acknowledgements
Jianpeng Liu thanks the financial support from the National Key R \& D program of China (grant no. 2024YFA1410400,  no. 2022YFA1604400/03 and no. 2020YFA0309601), the National Natural Science Foundation of China (Grant No. 12174257), the Science and Technology Commission of the Shanghai Municipality (Grant No. 21JC1405100), and the start-up grant of ShanghaiTech University. Xin Lu thanks the financial support from the National Natural Science Foundation of China (Grant no. 12404221),
\bibliography{references_review}

\end{document}